\documentclass[final,3p,times]{elsarticle}

\usepackage{graphicx}
\usepackage{amssymb}
\usepackage{amsmath}
\usepackage{amssymb}
\usepackage{amsthm}
\usepackage[utf8]{inputenc}
\usepackage[english]{babel}
\usepackage{xcolor}
\usepackage{threeparttable}
\usepackage{multirow}
\usepackage{graphicx}
\usepackage{algorithm}
\usepackage{algorithmic,setspace}
\usepackage{lipsum}
\usepackage{subfig}
\usepackage{caption}
\theoremstyle{remark}
\newtheorem*{remark}{Remark}                 

\journal{}

\begin{document}

\begin{frontmatter}


\title{Variational Probabilistic Multi-Hypothesis Tracking}

\author{Shuoyuan Xu, Hyo-Sang Shin and Antonios Tsourdos}
\ead{shuoyuan.xu@cranfield.ac.uk, h.shin@cranfield.ac.uk, a.tsourdos@cranfield.ac.uk}
\cortext[cor1]{Corresponding author: Hyo-Sang Shin}
\address{School of Aerospace, Transport and Manufacturing, Cranfield University, Cranfield MK43 0AL, UK}


\begin{abstract}
This paper proposes a novel multi-target tracking (MTT) algorithm for scenarios with arbitrary numbers of measurements per target. We propose the variational probabilistic multi-hypothesis tracking (VPMHT) algorithm based on the variational Bayesian expectation-maximisation (VBEM) algorithm to resolve the MTT problem in the classic PMHT algorithm. With the introduction of variational inference, the proposed VPMHT handles track-loss much better than the conventional probabilistic multi-hypothesis tracking (PMHT) while preserving a similar or even better tracking accuracy. Extensive numerical simulations are conducted to demonstrate the effectiveness of the proposed algorithm.
\end{abstract}


\end{frontmatter}




\section{Introduction}
{Multi-target tracking (MTT) refers to the technology of simultaneously estimating the number and states of targets given imperfect sensor measurements. MTT is a key technology for various applications for autonomous systems such as autonomous driving \cite{cho2014multi}, airborne surveillance \cite{kanade1998advances}, and pedestrians tracking \cite{scheunert2004multi} systems. Traditionally, multi-target tracking (MTT) algorithms are developed for scenarios with each target generating at most one measurement per scan. Scenarios like such are considered following the well-known "single measurement-to-target" assumption in MTT research. The most common MTT algorithms based on "single measurement-to-target" assumption include global nearest neighbour (GNN) \cite{konstantinova2003study}, multiple hypothesis tracking (MHT) \cite{blackman2004multiple, thomaidis2013multiple}, probabilistic data association (PDA) \cite{musicki1994integrated}, and Joint probabilistic data association (JPDA) \cite{hamid2015joint, he2020distributed}. There is a wide range of applications for trackers with the "single measurement-to-target" assumption. However, complicated scenarios such as MTT in sensor networks \cite{cho2014multi, kanade1998advances, scheunert2004multi} or extended object tracking \cite{feldmann2010tracking, gilholm2005spatial,baum2011shape, zhang2017box} often render the "single measurement-to-target" assumption invalid.

For MTT problems without the "single measurement-to-target" assumption, it is prevalent to use combinatory tracking approaches which incorporate multi-dimensional association (MDA) then fusion \cite{granstrom2019poisson, he2018multi}. Probabilistic multi-hypothesis tracking (PMHT) is one of the few methods that solve such MTT problems in a unified framework \cite{streit1995probabilistic}. The PMHT algorithm considers measurement-to-target association events independent of each other across all measurements. Multiple measurements are hence allowed to be assigned to one target. The independent modelling of measurement associations allows the complexity of PMHT to be linear in the number of targets, number of measurements, and time batch size. On the contrary, approaches based on the "single measurement-to-target" assumption often yields an NP-hard or NP-complete problem. Furthermore, the probabilistic framework of PMHT brings easy extensibility and lead to many variations of PMHT in different applications such as extended object tracking \cite{bordonaro2015extracting, bordonaro2017extended}, multiple sensors \cite{molnar1994application, krieg1997multisensor}, and target-manoeuvre \cite{logothetis1997maneuvering, ruan1998pmht}. However, PMHT generally performs poorly when track-loss occurs. More specifically, PMHT cannot sense an impending track-loss or an already lost target and tend to welcome several clutter measurements to already lost tracks \cite{Willett2002pmht}.

Many variations of PMHT have been developed to overcome the limitations when it comes to track-loss. Stand-alone track management systems are considered viable solutions for handling track-loss. Multiple hypothesis tracking (MHT) \cite{dunham2002hybrid} as add-on track management to PMHT shows significant improvement of the track-loss performance but with a high computational cost. The hypothesis test \cite{wieneke2007sequential, wieneke2008track} are also applied for track management in PMHT, which shares similar strengths and weaknesses of MHT. When it comes to improving the track-loss performance within the PMHT framework, the homothetic PMHT \cite{rago1995comparison, rago1995direct} is a well-applied method. Homothetic PMHT improves track-loss performance by using multiple measurement models with different noise covariances and the same mean for each target. However, the additional measurement noise covariances also lead to tracking accuracy degradation. The multi-model implementation in homothetic PMHT inspired many PMHT modifications such as multi-frame PMHT \cite{ruan1998pmht}, spirograph PMHT \cite{willett1998variety}, and adaptive homothetic PMHT \cite{jeong2002based}. Unfortunately, these multi-model PMHT approaches fail to show consistent improvements to homothetic PMHT. Apart from using multiple measurement models, the idea of applying more sophisticated measurement models is also exploited by implementing the measurement model of PDAF into PMHT \cite{rago1995modified}. This PMHT variant also shows little sign of track-loss handling capabilities. Clutter handling methods can partially reduce the effect of track-loss since they can identify the clutter measurements and prevent them from being assigned to lost targets. The detection-oriented PMHT \cite{Willett2002pmht} adds a clutter related constant in the denominator of the assignment probabilities, which shows promising results in clutter removal. The manoeuvring model implemented PMHT algorithms \cite{logothetis1997maneuvering, ruan1998pmht}, regardless of the target dynamics, are also strong performers for identifying clutters. Nonetheless, from the extensive simulation comparisons in \cite{willett1998variety, Willett2002pmht}, the existing variations of PMHT are not guaranteed to outperform even the simple PDAF when there are track-losses.

Inspired by the literature mentioned above, this paper aims to develop an MTT algorithm capable of handling track-loss while maintaining the strong points of PMHT. It is believed that the limited track-loss performance of PMHT is led by the expectation-maximisation (EM) algorithm applied, which requires the exact number of mixture components (number of targets in MTT) as prior knowledge \cite{lan2016survey, lundgren2015variational}. In variational Bayesian expectation-maximisation (VBEM), the number of mixture components can be set large, and through optimisation, the components that provide insufficient contribution describing the dataset will have their weight converging to zero \cite{Approximate_Inference}. Therefore, it is intuitive to utilise VBEM to handle the track-loss issue of PMHT. This paper develops a novel MTT algorithm referred to as the variational probabilistic multi-hypothesis tracking (VPMHT) by utilising the well-known VBEM to solve the MTT model of PMHT. The proposed algorithm offers an excellent track-loss handling capability and provides a better fusion accuracy than PMHT regardless of track-loss. Moreover, the proposed VPMHT is more robust against uncertainties like measurement noise and clutters. The performance of the proposed multi-sensor MTT algorithm is validated through extensive numerical simulations.}

The rest of the paper is organised as follows. Section II presents some preliminaries and backgrounds. Section III presents the classic PMHT algorithm and some analysis about the effect of track-loss in PMHT. Section IV derives the VPMHT algorithm on the basis of PMHT using the VBEM algorithm. In Section V, the performance analysis of the proposed VPMHT is done via theoretical derivations and numerical simulations. Finally, some discussions and conclusions are offered.

\section{Backgrounds and Preliminaries}

\subsection{System Model}
The set of target states and the set of measurements at the $t$th time step is denoted as:
\begin{equation}
\begin{aligned}
&\mathbf{X} = \{\mathbf{X}_0, \ldots, \mathbf{X}_t, \ldots, \mathbf{X}_T\}, \quad \mathbf{X}_t = \{ x_t^1, \ldots, x_t^i, \ldots, x_t^{N^{tar}}\}\\
&\mathbf{Z} = \{\mathbf{Z}_0, \ldots, \mathbf{Z}_t, \ldots, \mathbf{Z}_T\}, \quad \mathbf{Z}_t = \{z_t^{1}, \ldots, z_t^{j}, \ldots, z_t^{N_t^{mea}}\}
\end{aligned}
\label{state_measurement_set_vpmht}
\end{equation}where $N^{tar}$ is the initial number of targets, $x_t^i$ denote the state of the $i$th target at the $t$th time step, $N_t^{mea}$ is the number of measurements at the $t$th time step, $z_t^j$ denote the $j$th measurement at the $t$th time step. In this paper, $i$, $j$, and $t$ denote the {indices} of target, measurement, and time step, respectively.  It is worth noting that PMHT assumes a known and constant number of targets $N^{tar}$, which is the major limitation of PMHT-based algorithms and will be further discussed later in this paper (3.3). 

Suppose the following target dynamics and measurement model:
\begin{equation}
\begin{aligned}
&x_t^{i}=F_{t-1}^{i}x_{t-1}^{i}+w_{t-1}^{i} \\
&\left. z_t^{j} \right|_{j=i}=H_t^{j}x_t^{i}+v_t^{j}
\end{aligned}
\end{equation}where matrix $F_t^{i}$ and $H_t^{j}$ is the state transition model and the observation model. $w_t^{i}$ and $v_t^{j}$ are process and observation noises, which are both assumed to be zero-mean Gaussian noise with covariances $Q_t^{i}$ and $R_t^{j}$. The subscript $\left.(\cdot)\right|_{j=i}$ indicates that the $j$th measurement is originated from the $i$th target. 

For simplicity, the following assumptions, which are common in MTT, are made.

\textbf{Assumption 1.} The clutter distribution is modelled as a Poisson point process (PPP) where the number of clutters at each sensor follows Poisson and the locations of clutters are assumed to be uniformly distributed in the surveillance region.

\textbf{Assumption 2.} Each target can generate multiple measurements and each measurement can originate from at most one target. Each target-generated measurement is independent of each other and multiple measurements can originate from the same target.

\subsection{Problem Formulation}
The MTT problem is defined as joint estimation of the measurement origins and target states. According to the system model, the probability density functions (PDFs) of target transition $\varphi_t^{i}\left(x_t^{i} \mid x_{t-1}^{i}\right)$, measurements of targets $\zeta_{j}\left(z_t^{j} \mid x_t^{i}\right)$, and measurements of clutters can be written as:

\begin{equation}
\begin{aligned}
&\varphi_t^{i}\left(x_t^{i} \mid x_{t-1}^{i}\right) = \mathcal{N}\left(x_t^{i} \mid F_t^i x_{t-1}^{i}, Q_t^i\right)\\
&\left. \zeta_{j}\left(z_t^{j} \mid x_t^{i}\right)  \right|_{j=i} = \mathcal{N}\left(z_t^{j} \mid H_t^{j} x_t^{i}, R_t^{j}\right) \\
&\left. \zeta_{j}(z_t^{j})  \right|_{j=0} = \mathcal{U}(A_t,B_t) 
\end{aligned}
\label{dyna_and_meas_vpmht}
\end{equation} where $\mathcal{N}(x \mid \mu, \Sigma)$ denotes the Gaussian distribution of variable $x$ with mean $\mu$ and covariance $\Sigma$. $\mathcal{U}(A_t,B_t)$ denotes the uniform distribution with $A_t$ and $B_t$ as the bounds, and $A_t-B_t$ is a constant representing the interval. $j=0$ represents that the measurement $j$ is originate from background clutters. The PDFs of target initials $x_0^i$ are given as prior knowledge for notational convenience:
\begin{equation}
\varphi_{0}^i(x_0^i)=\mathcal{N}\left(x_0^i \mid \bar{x}_0^i, \bar{\Sigma}_0^i\right)
\end{equation}where the mean $\bar{x}_0^i$ and covariance matrix $\bar{P}_0^i$ are the initial state estimate and corresponding error covariance of target $i$. For the convenience of later derivations, we define the measurement-to-target association indicators as:
\begin{equation}
\mathbf{L}_t=\left\{L_t^{1}, \ldots, L_t^{j}, \ldots, L_t^{N_t^{mes}}\right\} 
\label{hidden_variables_vpmht}
\end{equation}where each $L_t^j$ is a vector 
\begin{equation}
L_t^j = (l_t^{j,1}, \ldots, l_t^{j,i}, \ldots, l_t^{j, N^{tar}}, l_t^{j, N^{tar}+1})^T.
\label{hidden_variables_each_vpmht}
\end{equation}The value $l_t^{j,i}$ satisfies $l_t^{j,i} \in\{0,1\}$ and $\sum_{i=1} l_t^{j,i}=1$. If $l_t^{j,i}=1$ and $i \leqslant N^{tar}$, it means that the $j$th measurement is from the $i$th target whereas $l_t^{j,i}=1$ and $i = N^{tar}+1$ indicates the $j$th measurement is from clutters. 

The goal of the VPMHT algorithm is to solve the joint parameter estimation problem of target states $\mathbf{X}$ and measurement-to-track association indicator $\mathbf{L}$. The likelihood function of parameters $\mathbf{X}$ and $\mathbf{L}$ based on all measurements $\mathbf{Z}$ is given as:
\begin{equation}
\begin{aligned}
{p(\mathbf{Z} \mid \mathbf{X}, \mathbf{L})} 
&= \prod_{t=1}^{T} p(\mathbf{Z}_t \mid \mathbf{X}_t, \mathbf{L}_t) \\ 
&= \prod_{t=1}^{T} \prod_{j=1}^{N_t^{mes}} p(z_t^{j} \mid \mathbf{X}_t, {L}_t^j) \\ 
&= \prod_{t=1}^{T} \prod_{j=1}^{N_t^{mes}} \left\{\mathcal{U}(A_t,B_t)^{l_t^{j, N^{tar}+1}}\prod_{i=1}^{N^{tar}} \zeta_{j}\left(z_t^{j} \mid x_t^{i}\right)^{l_t^{j,i}}\right\}.  
\end{aligned}
\label{objective_vpmht}
\end{equation} By solving the joint parameter estimation problem of $\mathbf{L}$ and $\boldsymbol{X}$ in this likelihood function, the MTT problem can be resolved. The difficulty of solving this likelihood function is that the association indicators $\mathbf{L}$ are hidden variables and coexist with model parameters $\boldsymbol{X}$. One way of solving this joint parameter estimation problem is using EM algorithm \cite{bilmes1998gentle} to iterativly find the maximum a posterior (MAP) estimates of the model parameters and hidden variables. The proposed VPMHT algorithm and the baseline algorithm PMHT are all developed based on EM algorithm.

\section{Probabilistic Multi-Hypothesis Tracking and Analysis}
In this section, we first introduce a mixture modelling of the MTT measurements as one of the premises in formulating PMHT and VPMHT. Then a brief review of the original PMHT algorithm is provided. Finally, we identify and discuss some practical issues of PMHT algorithms.

\subsection{Mixture PDF of Measurements}
EM algorithm in PMHT iteratively finds maximum a posterior (MAP) estimates of target states and association vectors. The target states are considered model parameters, and association indicators are considered hidden variables in PMHT. In a classic EM algorithm architecture, the expectation (E) step computes the expectations of the hidden variables using the current estimate for the parameters. Then, the maximisation (M) step maximises the parameters with the expected hidden variables from E-step \cite{bilmes1998gentle}. It can be identified that a connection between hidden variables and model parameters is required to perform the E-step. In PMHT, such a connection is developed through the introduction of a mixture model among the measurements.

To formulate the EM framework of PMHT, the prior probability of a measurement spawns from target $i$, ${\pi}_t^i$ is introduced to link the model parameters, and hidden variables \cite{bordonaro2015extracting, molnar1994application, Willett2002pmht}. Apart from its mathematical usage in deriving our algorithm, the value $\pi$ is physically critical since some targets may produce more measurements per scan than others because of individual target characteristics (e.g., signal-to-noise ratio), environmental effects, sensor properties, or other application considerations. A mixture model around the measurements $\mathbf{Z}$ and target states $\mathbf{X}$ can be formulated as:

\begin{equation}
\begin{aligned}
{p(\mathbf{Z} \mid \mathbf{X}, \boldsymbol{\pi})} 
&= \prod_{t=1}^{T} p(\mathbf{Z}_t \mid \mathbf{X}_t, \boldsymbol{\pi}_t) \\ 
&= \prod_{t=1}^{T} \prod_{j=1}^{N_t^{mes}} p(z_t^{j} \mid \mathbf{X}_t, \boldsymbol{\pi}_t) \\
&= \prod_{t=1}^{T} \prod_{j=1}^{N_t^{mes}} \left\{ \pi_t^{N^{tar}+1}\mathcal{U}(A_t,B_t) + \sum_{i=1}^{N^{tar}} \pi_t^i \zeta_{j}\left(z_t^{j} \mid x_t^{i}\right)\right\}
\end{aligned}
\label{mixture_model}
\end{equation} here, the set of prior probability of measurement to target association can be formulated as:
\begin{equation}
\begin{aligned}
\boldsymbol{\pi} = \{\boldsymbol{\pi}_0, \ldots, \boldsymbol{\pi}_t, \ldots, \boldsymbol{\pi}_T\}, \quad \boldsymbol{\pi}_t = \{\pi_t^1, \ldots, \pi_t^i, \ldots, \pi_t^{N^{tar}+1}\}
\end{aligned}
\end{equation}where $\sum_{i=1}^{N^{tar}+1} \pi_t^{i}=1$ and $0 \leqslant \pi_t^{i} \leqslant 1 \label{GMM_cons_vpmht}$. This mixture model implies that each measurement is generated by the weighted sum of all candidate targets. Similar mixture modelling is quite common in classic MTT algorithms \cite{he2020information} where each target's posterior is assumed as a mixture model of all possible associations. Here we make another common PMHT assumption to simplify the derivation:

\textbf{Assumption 3.} Measurements $\mathbf{Z}$ and association related parameters $\mathbf{L}$ and $\boldsymbol{\pi} $ are time independent:
\begin{equation}
\begin{aligned}
&p(\mathbf{L}_t \mid \mathbf{L}_{t-1}) = p(\mathbf{L}_t) \\
&p(\boldsymbol{\pi}_t \mid \boldsymbol{\pi}_{t-1}) = p(\boldsymbol{\pi}_t)\\
&p(\mathbf{Z}_t \mid \mathbf{Z}_{t-1}) = p(\mathbf{Z}_t) \\
\end{aligned}
\end{equation}

\subsection{Probabilistic Multi-Hypothesis Tracking}

The original Probabilistic Multi-Hypothesis Tracking (PMHT) algorithm, presented in \cite{streit1995probabilistic}, gives a probabilistic approach to solve the measurement-to-track assignment problem. Instead of strictly assigning each measurement to a track or clutter, PMHT estimates the association of measurements to tracks within a probabilistic framework using the expectation-maximisation (EM) method. There are no constraints imposed on the number of measurements originated per target in each scan. With an independent assumption across the measurements on the association process, PMHT avoids the combinatorial complexity lead by enumerating all association hypotheses, thus, yields a tractable complexity.

The solution of the PMHT algorithm is obtained through three interrelated EM steps: expectation step (E-step), forward maximisation step (Forward M-step), and backward maximisation step (Backward M-step). E-step computes the conditional density of the discrete association variable $L$ using Bayes'theorem given the measurements and state estimates computed in the last iteration. Forward M-step marginalises the batch joint PDF of target states over the measurements and associations, resulting in the target states estimates. Backward M-step further corrects the target state estimates at the current scan using target state estimates of future time steps with a smoother-like approach. Take a linear Gaussian system of Equation (\ref{objective_vpmht}) and (\ref{mixture_model}) as an example, the PMHT algorithm (see \textbf{Algorithm 1} for pseudo-code) at each time step is formulated as:

\begin{enumerate}
\item Initialisation of the target state estimates and their corresponding covariances using the estimates from the last time step $\{ x_{t-1| t-1}^1, ..., x_{t-1| t-1}^{i}, ...,  x_{t-1| t-1}^{N^{tar}}\}$ and $\{ P_{t-1| t-1}^1, ..., P_{t-1| t-1}^{i}, ...,  P_{t-1| t-1}^{N^{tar}}\}$, where ${x}_{t|t}^i$ and ${P}_{t|t}^i$ are the a posterior estimate and its corresponding error covariance matrix of target $i$ at time $t$.

\item E-step: Computing the association probabilities $w_{t}^{i,j}$ for all targets and clutters $i=1,2, \ldots, N^{tar}+1$, all time steps in the current batch $t=t, t+1, \ldots, t+t_{batch}$, as well as all measurements $j=1,2, \ldots, N_{meas}$, according to

\begin{equation}
w_{t}^{i,j} = \frac{\pi_t^{i} \mathcal{N}\left(z_t^{j} \mid H_t^{j} x_{t \mid t}^{i}, R_t^{j}\right)}{\sum_{i=1}^{N^{tar}}\left[\pi_t^{i} \mathcal{N}\left(z_t^{j} \mid H_t^{j} x_{t \mid t}^{i}, R_t^{j}\right)\right]}
\end{equation} The mixture weight for each target at each scan can be updated as:
\begin{equation}
\pi_t^{i} = \frac{1}{N^{tar}} \sum_{i=1}^{N^{tar}} w_{t}^{i,j}
\end{equation}

\item Forward M-step: Computing the synthetic measurements $\tilde{z}_{t}^{i}$ and their associated (synthetic) measurement covariances $\tilde{R}_{t}^{i}$ for each target $i$:
\begin{equation}
\tilde{z}_{t}^{i} = \frac{1}{n_{t} \pi_t^{i}} \sum_{r=1}^{n_{t}} w_{t}^{i,j} z_t^{j}
\; \;  \text{and} \; \; 
\tilde{R}_{t}^{i} = \left[n_{t} \pi_t^{i}\right]^{-1} R_{t}^{j}
\end{equation} For each target $i$, a standard Kalman filter algorithm \cite{kalman1960new} is applied to obtain the estimates of each trajectory $x_{t \mid t}^{i}$ and its corresponding covariance $P_{t \mid t}^{i}$ with the synthetic measurements and corresponding covariances $\tilde{z}_{t}^{i}$ and $\tilde{R}_{t}^{i}$: 

\begin{equation}
\begin{aligned}
&x_{t \mid t-1}^i =F_{t}^{i} x_{t-1 \mid t-1}^i \\
&P_{t \mid t-1}^i =F_{t}^{i} P_{t-1 \mid t-1}^i\left(F_{t}^{i}\right)^{T}+Q_{t}^{i} \\
&K_{t}^i =P_{t \mid t-1}^i\left(H_{t}^{i}\right)^{T}\left\{H_{t}^{i} P_{t \mid t-1}^i\left(H_{t}^{i}\right)^{T}+\tilde{R}_{t}^{i}\right\}^{-1} \\
&x_{t \mid t}^i = x_{t \mid t-1}^i + K_{t}^i\left(\tilde{z}_{t}^{i} - H_t^i x_{t \mid t-1}^i\right) \\ 
&P_{t \mid t}^i =P_{t \mid t-1}^i-K_{t}^i H_{t}^{i} P_{t \mid t-1}^i \\
\end{aligned}
\end{equation} where ${x}_{t|t-1}^i$ and ${P}_{t|t-1}^i$ are the prior state estimate and its corresponding error covariance matrix of target $i$ at time $t$, and $K_{t}^i$ is the Kalman gain of target $i$ at time $t$ .

\item Backward M-step: For each target $i=1,2, \ldots, {N^{tar}}$, Kalman smoothing algorithm \cite{rauch1965maximum} is applied to obtain the smoothed estimates $\hat{x}_{t \mid t}^i$ and their corresponding error covariances $\hat{P}_{t \mid t}^i$:

\begin{equation}
\begin{aligned}
\hat{x}_{t \mid t}^i &= x_{t \mid t}^i+P_{t \mid t}^i\left(F_{t+1}^{i}\right)^{T} (P_{t \mid t-1}^i)^{-1}\left[\hat{x}_{t+1 \mid t+1}^i-F_{t+1}^{i} x_{t \mid t}^i\right]\\
\hat{P}_{t \mid t}^i &=P_{t \mid t}^i+P_{t \mid t}^i\left(F_{t+1}^{i}\right)^{T} (P_{t \mid t-1}^i)^{-1} \left[\hat{P}_{t+1 \mid t+1}^i-P_{t \mid t-1}^i\right] (P_{t \mid t-1}^i)^{-1} \left(F_{t+1}^{i}\right)^{T} P_{t \mid t}^i
\end{aligned}
\end{equation}

\item Return to step 2 for the next EM iteration unless a stopping criterion is reached. One of the simplest approaches to end the computation is to set a fixed number of iterations \cite{rago1995comparison, Willett2002pmht}. To improve the computational efficiency the PMHT likelihood function (Equation (\ref{mixture_model})) is applied as the stop criterion \cite{Willett2002pmht}. The EM iteration is set to stop when the change of the criterion value falls below a given level.
\end{enumerate}

\begin{algorithm*}
\caption{PMHT algorithm at scan $t$}
\begin{algorithmic}[1]
\renewcommand{\algorithmicrequire}{\textbf{Input:}}
\renewcommand{\algorithmicensure}{\textbf{Output:}}
\REQUIRE The bounds of the clutter distribution $(A_t,B_t)$, number of targets $N^{tar}$, the smoothed previous estimation $\left\{\hat{x}_{t-1 \mid t-1}^{i}, \hat{P}_{t-1 \mid t-1}^{i}\right\},$ received measurements $\mathbf{Z}$, and termination condition $\bigtriangleup \mathcal{L}_{terminate}$
\ENSURE  The smoothed current estimation $\{\hat{x}_{t \mid t}^i, \hat{P}_{t \mid t}^i\}$
\WHILE{$\bigtriangleup \mathcal{L} > \bigtriangleup \mathcal{L}_{terminate}$}
\FOR{$t = t, \ldots, T$}
\STATE Initialisation \\ \quad
\begin{math}
\begin{aligned}
& \{ x_{t-1| t-1}^1, ..., x_{t-1| t-1}^{i}, ...,  x_{t-1| t-1}^{N^{tar}}\} \\
& \{ P_{t-1| t-1}^1, ..., P_{t-1| t-1}^{i}, ...,  P_{t-1| t-1}^{N^{tar}}\} \\
\end{aligned}
\end{math}
\STATE E-step \\
Computing the association probabilities $w_{t}^{i,j}$ for all targets $i$ and all measurements $j$\\
\quad
\begin{math}
\begin{aligned}
w_{t}^{i,j} &= \frac{\pi_t^{i} \mathcal{N}\left(z_t^{j} \mid H_t^{j} x_{t \mid t}^{i}, R_t^{j}\right)}{\sum_{i=1}^{N^{tar}}\left[\pi_t^{i} \mathcal{N}\left(z_t^{j} \mid H_t^{j} x_{t \mid t}^{i}, R_t^{j}\right)\right]} 
\end{aligned}
\end{math} \\
Update the mixture weights\\
\quad
\begin{math}
\begin{aligned}
\pi_t^{i} = \frac{1}{N^{tar}} \sum_{i=1}^{N^{tar}} w_{t}^{i,j}
\end{aligned}
\end{math}\\

\STATE Forward M-step \\
Computing the synthetic measurements $\tilde{z}_{t}^{i}$ and their corresponding error covariances $\tilde{R}_{t}^{i}$ for each target $i$ \\
\quad
\begin{math}
\begin{aligned}
&\tilde{z}_{t}^{i} = \frac{1}{n_{t} \pi_t^{i}} \sum_{r=1}^{n_{t}} w_{t}^{i,j} z_t^{j} \\
&\tilde{R}_{t}^{i} = \left[n_{t} \pi_t^{i}\right]^{-1} R_{t}^{j}
\end{aligned}
\end{math} \\
Updating the state estimate of each target $x_{t \mid t}^{i}$ and its covariance $P_{t \mid t}^{i}$ using Kalman filter \\
\quad
\begin{math}
\begin{aligned}
&x_{t \mid t-1}^i =F_{t}^{i} x_{t-1 \mid t-1}^i \\
&P_{t \mid t-1}^i =F_{t}^{i} P_{t-1 \mid t-1}^i\left(F_{t}^{i}\right)^{T}+Q_{t}^{i} \\
&K_{t}^i =P_{t \mid t-1}^i\left(H_{t}^{i}\right)^{T}\left\{H_{t}^{i} P_{t \mid t-1}^i\left(H_{t}^{i}\right)^{T}+\tilde{R}_{t}^{i}\right\}^{-1} \\
&x_{t \mid t}^i = x_{t \mid t-1}^i + K_{t}^i\left(\tilde{z}_{t}^{i} - H_t^i x_{t \mid t-1}^i\right) \\ 
&P_{t \mid t}^i =P_{t \mid t-1}^i-K_{t}^i H_{t}^{i} P_{t \mid t-1}^i
\end{aligned}
\end{math} \\
\ENDFOR
\FOR{$t = T, \ldots, t$}
\STATE Backward M-step \\
Computing smoothed variational parameters ${\pi}_t^i$, $\hat{x}_{t \mid t}^i$, and $\hat{P}_{t \mid t}^i$ \\ \quad
\begin{math}
\begin{aligned}
\hat{x}_{t \mid t}^i &= x_{t \mid t}^i+P_{t \mid t}^i\left(F_{t+1}^{i}\right)^{T} (P_{t \mid t-1}^i)^{-1}\left[\hat{x}_{t+1 \mid t+1}^i-F_{t+1}^{i} x_{t \mid t}^i\right]\\
\hat{P}_{t \mid t}^i &=P_{t \mid t}^i+P_{t \mid t}^i\left(F_{t+1}^{i}\right)^{T} (P_{t \mid t-1}^i)^{-1} \left[\hat{P}_{t+1 \mid t+1}^i-P_{t \mid t-1}^i\right] (P_{t \mid t-1}^i)^{-1} \left(F_{t+1}^{i}\right)^{T} P_{t \mid t}^i
\end{aligned}
\end{math}
\ENDFOR
\STATE Computing the PMHT likelihood as the stop criterion\\ \quad
\begin{math} 
\begin{aligned}
{p(\mathbf{Z} \mid \mathbf{X}, \boldsymbol{\pi})} = \prod_{t=1}^{T} \prod_{j=1}^{N_t^{mes}} \left\{ \pi_t^{N^{tar}+1}\mathcal{U}(A_t,B_t) + \sum_{i=1}^{N^{tar}} \pi_t^i \zeta_{j}\left(z_t^{j} \mid x_t^{i}\right)\right\}
\end{aligned}
\end{math}
\ENDWHILE
\end{algorithmic}
\end{algorithm*}

\subsection{Some Practical Issues of PMHT}

From the Linear-Gaussian PMHT algorithm presented in the previous section, some practical issues can be identified and will be discussed in this section. We start the analysis with an assumption made in PMHT derivation that a fixed number of targets is known to the model. Then the discussions of the synthetic measurements in the M-step are followed.

The original PMHT algorithm \cite{streit1995probabilistic} assumes a fixed and known number of targets $N^{tar}$. Since PMHT is an EM-based algorithm, it does not allow for a change of cardinality (number of targets) \cite{Approximate_Inference}. Thus PMHT has no means of track initiation or tracks deletion \cite{springer2013mathematical}. In practice, the lack of track deletion capabilities of PMHT often deteriorates the tracking performance by assigning measurements from other targets or clutters to already lost tracks. Handling such incapabilities are extremely challenging. Even add-on track management systems can not guarantee to eliminate the effects of track-losses for PMHT algorithms \cite{Willett2002pmht}. Implementing the functionality of handling track-loss scenarios to PMHT-like algorithms is one of the main focuses of this paper. 

The synthetic measurements applied in the M-step is another problem to be addressed. The synthetic measurements and their corresponding error covariances are computed through the direct weighted sum of all measurements and error covariance with association probabilities as the weights. Such weighted sum computation introduces several issues to the PMHT algorithm. First, the direct weighted sum of measurements requires all measurements to be composed of identical physical quantities, such as positions in the same coordinate system. The Identical -physical-quantity requirement limits the usage of heterogeneous sensors or requires extra treatments of the measurements, e.g. coordinate transformation. Secondly, the weighted sum computation does not consider the error covariances, which implies that the accuracy levels of all sensors are considered the same. For various multi-sensor systems with mixed quality sensors \cite{he2020distributed}, ignoring the accuracy differences among sensors could be impractical. Finally, from the basic properties of Gaussian distributions \cite{hogg2005introduction}, the calculation of error covariance of the synthetic measurements are mere approximations, which does not fully represent the uncertainties of the synthetic measurements.

\newpage
\section{Variational Probabilistic Multi-Hypothesis Tracking}
The variational Bayesian expectation-maximisation (VBEM) is a classical method of finding the MAP distribution of all the model parameters and hidden variables. VBEM algorithm tackles the challenging joint parameter estimation problems (Equation (\ref{objective_vpmht}) and (\ref{mixture_model})) through iteratively optimising a variational joint distribution of hidden variables and model parameters. 

\subsection{Objective of VPMHT}
Our proposed VPMHT algorithm aims to find a variational distribution $q(\mathbf{L}, \boldsymbol{\pi}, \mathbf{X})$ that best approximates the posterior distribution $p(\mathbf{L}, \boldsymbol{\pi}, \mathbf{X} \mid \mathbf{Z})$. From variational inference theory \cite{Approximate_Inference}, the best approximated $q(\mathbf{L}, \boldsymbol{\pi}, \mathbf{X})$ is found by minimising the Kullback-Leibler (KL) divergence :

\begin{equation}
\mathrm{KL}(q \| p) =-\int q(\mathbf{L}, \boldsymbol{\pi}, \mathbf{X}) \ln \left\{\frac{p(\mathbf{L}, \boldsymbol{\pi}, \mathbf{X} \mid \mathbf{Z})}{q(\mathbf{L}, \boldsymbol{\pi}, \mathbf{X})}\right\} \mathrm{d} \mathbf{L} \mathrm{d} \boldsymbol{\pi} \mathrm{d} \mathbf{X}.
\end{equation} We assume the variational distribution $q(\mathbf{L}, \boldsymbol{\pi}, \mathbf{X})$ can be decomposed by the mean-field approximation:
\begin{equation}
q(\mathbf{L}, \boldsymbol{\pi}, \mathbf{X})=q(\mathbf{L}) q(\boldsymbol{\pi}, \mathbf{X}),
\label{variational_distribution_vpmht}
\end{equation} the conditions $q^{\star}(\mathbf{L})$ and $q^{\star}(\boldsymbol{\pi}, \mathbf{X})$ that minimise KL divergence \cite{Approximate_Inference} are:
\begin{equation}
\ln q^{\star}(\mathbf{L}) = \mathbb{E}_{\boldsymbol{\pi}, \mathbf{X}}[\ln  p(\mathbf{Z}, \mathbf{L}, \boldsymbol{\pi}, \mathbf{X})] + \text {const}
\label{q*z_vpmht}
\end{equation}
\begin{equation}
\ln q^{\star}(\boldsymbol{\pi}, \mathbf{X}) = \mathbb{E}_{\mathbf{L}}[\ln p(\mathbf{Z}, \mathbf{L}, \boldsymbol{\pi}, \mathbf{X})] + \text {const}.
\label{q*pi_mu_A_vpmht}
\end{equation} 

It can be identified that the joint distribution $p(\mathbf{Z}, \mathbf{L}, \boldsymbol{\pi}, \mathbf{X})$ is crucial to the compute both $q^{\star}(\mathbf{L})$ and $q^{\star}(\boldsymbol{\pi}, \mathbf{X})$. We calculate $p(\mathbf{Z}, \mathbf{L}, \boldsymbol{\pi}, \mathbf{X})$ by decomposing it using Bayesian theorem:
\begin{equation}
\begin{aligned}
&p(\mathbf{Z}, \mathbf{L}, \boldsymbol{\pi}, \mathbf{X})\\
&=p(\mathbf{Z} | \mathbf{L}, \mathbf{X}) p(\mathbf{L} | \boldsymbol{\pi}) p(\boldsymbol{\pi}) p(\mathbf{X})
\end{aligned}
\label{joint_decompose_vpmht}
\end{equation}where $p(\mathbf{Z} | \mathbf{L}, \mathbf{X})$ can be obtained according to the problem formulation (\ref{objective_vpmht}), the conditional distribution $p(\mathbf{L} | \boldsymbol{\pi})$ can be written as
\begin{equation}
p(\mathbf{L} | \boldsymbol{\pi}) = \prod_{t=1}^{T} \prod_{j=1}^{N^{mes}_t} \prod_{i=1}^{{N^{tar}+1}} (\pi^i_t)^{l^{j,i}_t} \text{.}
\label{vb_z|pi_vpmht}
\end{equation} The priors of targets, using the transient model (\ref{dyna_and_meas_vpmht}), can be written as:
\begin{equation}
p(\mathbf{X}) = \prod_{i=1}^{{N^{tar}}} \varphi_{0}^i(x_0^i) \left\{ \prod_{t=1}^{T} \varphi_t^{i} (x_t^{i} \mid x_{t-1}^{i}) \right\} \text{.}
\label{X_prior}
\end{equation} It can be identified that the prior distribution of each target at each time step follows Gaussian, which is a conjugate distribution. The value of $p(\boldsymbol{\pi})$ is set via choosing the prior of the mixing coefficient. A conjugate prior, Dirichlet, over the mixing coefficients $\boldsymbol{\pi}$ is selected for a straightforward derivation:
\begin{equation}
p(\boldsymbol{\pi})=\prod_{t=1}^{T} \operatorname{Dir}\left(\boldsymbol{\pi}_t | \boldsymbol{\alpha}_{0}\right)= \prod_{t=1}^{T} \left\{ C\left(\boldsymbol{\alpha}_{0}\right) \prod_{i=1}^{{N^{tar}+1}} (\pi_t^i)^{\alpha_{0}-1}\right\} 
\label{VGMM_pi_prior_vpmht}
\end{equation}where $C\left(\boldsymbol{\alpha}_{0}\right)$ is the normalisation constant, parameter $\alpha_0$ is the concentration parameters. By substituting Equations (\ref{objective_vpmht}), (\ref{vb_z|pi_vpmht}), (\ref{X_prior}), and (\ref{VGMM_pi_prior_vpmht}) into Equation (\ref{joint_decompose_vpmht}), the joint distribution $\ln p(\mathbf{Z}, \mathbf{L}, \boldsymbol{\pi}, \mathbf{X})$ can be rewritten as:

\begin{equation}
\begin{aligned}
&\ln{p(\mathbf{Z}, \mathbf{X}, \mathbf{L}, \boldsymbol{\pi})} \\
&= \sum_{i=1}^{N^{tar}} \left(\varphi_0^{i}\left(x_{0}^{i}\right) + \sum_{t=1}^T \left[ \ln \varphi_t^{i}\left(x_t^{i} \mid x_{t-1}^{i}\right) + \sum_{j=1}^{N^{mes}_t} (l_t^{j, i} \ln{ \pi_t^i}) \right. \right.\\
& + \left. \left. \sum_{j=1}^{N^{mes}_t} [l_t^{j, i} \ln{\zeta_{j}\left(z_t^{j} \mid x_t^{i}\right)}] \right] \right)  + \sum_{t=1}^T \sum_{j=1}^{N^{mes}_t} {l_t^{j, N^{tar}+1}} \ln \mathcal{U}(A_t,B_t) + \sum_{t=1}^T \ln{\operatorname{Dir}\left(\boldsymbol{\pi}_t | \boldsymbol{\alpha}_{0}\right)}
\end{aligned}
\label{derived_joint_distribution_vpmht}
\end{equation}

\subsection{Variational Distribution of Hidden Variables}
\label{responsibility}
Substituting Equation (\ref{derived_joint_distribution_vpmht}) into (\ref{q*z_vpmht}), the general expression for the solution of condition $q^{*}(\mathbf{L})$ can be obtained as: \begin{equation}
\begin{aligned}
\ln q^{\star}(\mathbf{L}) &= \mathbb{E}_{\boldsymbol{\pi}, \mathbf{X}}[\ln{P(\mathbf{Z}, \mathbf{X}, \mathbf{L}, \boldsymbol{\pi})}] + \text {const}\\
&=\sum_{t=1}^T \sum_{i=1}^{N^{tar}} \sum_{j=1}^{N^{mes}_t} \left( \mathbb{E}_{\boldsymbol{\pi}}[\ln{ \pi_t^i}] - \frac{1}{2} \mathbb{E}_{\mathbf{X}}[(H_t^j x_t^i - z_t^j)^{T}(R_t^j)^{-1}(H_t^j x_t^i - z_t^j)] - \frac{D_t^j}{2}\ln{2\pi} -  \frac{D_t^j}{2}\ln{\operatorname{det}(R_t^j)}\right) \\
&+ \sum_{t=1}^T \sum_{j=1}^{N^{mes}_t} l_t^{j, N^{tar}+1} (\frac{\pi_t^{N^{tar}+1}}{B_t-A_t}) + \text {const}.
\end{aligned}
\label{q*l}
\end{equation} By setting 
\begin{equation}
\ln \rho_t^{j,i}=\left\{\begin{array}{ll}
\mathbb{E}_{\boldsymbol{\pi}}[\ln{ \pi_t^i}] -\frac{1}{2} \mathbb{E}_{\mathbf{X}}[(H_t^j x_t^i - z_t^j)^{T}(R_t^j)^{-1}(H_t^j x_t^i - z_t^j)] \\ 
- \frac{D_t^j}{2}\ln{2\pi} -  \frac{D_t^j}{2}\ln{\operatorname{det}(R_t^j)} & \text { if } i \in \{1, \ldots, N^{tar}\} \\
\frac{\pi_t^{N^{tar}+1}}{B-A} & \text { if } i =N^{tar}+1
\end{array}\right.
\label{rho_vpmht}
\end{equation} 
\begin{equation}
\begin{aligned}
r_t^{j,i} &= \frac{\rho_t^{j, i}}{\sum_{i=1}^{N^{tar}} \rho_t^{j, i}}
\end{aligned}
\end{equation} $\ln q^{\star}(\mathbf{L}_t)$ can be simplified as:
\begin{equation}
q^{\star}(\mathbf{L}_t)=\prod_{t=1}^{T} \prod_{j=1}^{N^{mes}_t} \prod_{i=1}^{{N^{tar}+1}} (r_t^{j,i})^{l_t^{j,i}}
\label{q_starz_wrt_r_vpmht}
\end{equation}where $D_t^j$ is the dimension of measurement ${z}_t^{j}$. It can be identified that $r_t^{j,i}$ is a good representation of $q^{\star}(\mathbf{L_t})$. Moreover, from the definition of $r^{j,i}$, the relationship between $r^{j,i}$ and $l^{j,i}$ can be established as $\mathbb{E} [l_t^{j,i}] = r_t^{j,i}$ and $r_t^{j,i}$ will be non-negative and always sum up to one. Therefore, $r_t^{j,i}$ can be considered as the representative confidence for the association of measurement $z_t^{j}$, to the $i$th target. Using the classical Kalman filter, we have:
\begin{equation}
\begin{aligned}
x_{t\mid t-1}^{i} &= F_t^i x_{t-1\mid t-1}^{i} \\
P_{t\mid t-1}^{i} &= F_t^i P_{t-1\mid t-1}^{i} (F_t^i)^{T} + Q_t^i 
\end{aligned}
\end{equation} Due to the conjugate properties of $\boldsymbol{\pi}$ and $\mathbf{X}$, the terms in Equation (\ref{rho_vpmht}) can be easily derived using the general properties of Dirichlet and Gaussian distributions respectively.

\begin{equation}
\begin{aligned}
&\mathbb{E}_{\boldsymbol{\pi}}\left[\ln \pi_t^{i}\right] = \psi\left(\alpha_t^i\right)-\psi(\widehat{\alpha}_t)\\
&\mathbb{E}_{\mathbf{X}}[(H_t^j x_t^i - z_t^j)^{T}(R_t^j)^{-1}(H_t^j x_t^i - z_t^j)]& \\
\quad \quad \quad &= \left(H_t^j x_{t \mid t}^i-z_t^j\right)^{\mathrm{T}}(R_t^j)^{-1}\left(H_t^j x_{t \mid t}^i-z_t^j\right)+ \operatorname{Tr}((H_t^j)^T (R_t^j)^{-1} H_t^j P_{t \mid t}^{i})
\end{aligned}
\label{epimua_vpmht}
\end{equation}where $\psi(\cdot)$ is the digamma function $\psi(\alpha) = \frac{d}{d \alpha} \ln \Gamma(\alpha)$, $\Gamma(\alpha)=(\alpha-1) !$, and $\widehat{\alpha_t}=\sum\alpha_t^{i}$.

\subsection{Variational Distribution of Model Parameters}
\label{model_parameters}
Similar to the derivation of $q^{\star}(\mathbf{L})$, we substitute Equation (\ref{derived_joint_distribution_vpmht}) into (\ref{q*pi_mu_A_vpmht}), yielding 
\begin{equation}
\begin{aligned}
\ln q^{\star}(\boldsymbol{\pi}, \mathbf{X}) &= \mathbb{E}_{\mathbf{L}}[\ln{p(\mathbf{Z}, \mathbf{X}, \mathbf{L}, \boldsymbol{\pi})}] + \text {const}\\
&= \sum_{t=1}^T \ln p(\boldsymbol{\pi}_t) +\sum_{i=1}^{N^{tar}} \ln \varphi_0^{i}\left(x_{0}^{i}\right) + \sum_{t=1}^T \mathbb{E}_{\mathbf{L}}[\ln p(\mathbf{L}_t \mid \boldsymbol{\pi}_t)] \\
&+ \sum_{t=1}^T \sum_{i=1}^{N^{tar}} \left\{ \varphi_t^{i}\left(x_t^{i} \mid x_{t-1}^{i}\right) + \sum_{j=1}^{N^{mes}_t} \left( \mathbb{E}_{\mathbf{L}}[l_t^{j,i}] \ln \zeta_{j}(z_t^{j} \mid x_t^{i})\right)\right\} + \text {const}
\label{q*pi_mu_A_expended}
\end{aligned}
\end{equation}

It can be seen from Equation (\ref{q*pi_mu_A_expended}) that terms containing $\boldsymbol{\pi}$ and $\mathbf{X}$ are independent to each other. This implies that the variational distribution of $\ln q^{\star}(\boldsymbol{\pi}, \mathbf{X})$ can be decomposes as:
\begin{equation}
\begin{aligned}
q(\boldsymbol{\pi}, \mathbf{X}) &= q(\boldsymbol{\pi}) q(\mathbf{X}) \\
&= \prod_{t=1}^T \left\{q(\boldsymbol{\pi}_t) \prod_{i=1}^{N^{tar}} q(x_i)\right\}
\end{aligned}
\end{equation}where \begin{equation}
\begin{aligned}
\ln q^{\star}(\boldsymbol{\pi}_t) &= \sum_{j=1}^{N^{mes}_t} \sum_{i=1}^{N^{tar}} r_t^{j, i} \ln \pi_t^{i} + \left(\alpha_{0}-1\right) \sum_{i=1}^{N^{tar}} \ln \pi_t^{i}  + \mathrm{const} \text{.}
\end{aligned}
\label{vari_pi_dir}
\end{equation} From the conjugate priors we defined, $q^{\star}(\boldsymbol{\pi})$ should be a Dirichlet distribution and the parameters can be easily obtained by reorganising terms in (\ref{vari_pi_dir}):
\begin{equation}
\alpha_t^i = \alpha_0 + \sum_{j=1}^{N^{mes}_t} r_t^{j,i}
\end{equation}

\subsubsection{Forward Filtering}
As for the variational distribution of target states, we model it through a two-step process of forward update then backward smoothing. The forward update leverage the Markov assumption which is commonly used in filter theories that the current target states are only conditionally dependent on the immediately previous states and current measurements:

\begin{equation}
\begin{aligned}
\ln q^{\star}(x_t^i)_{forward} &= \mathbb{E}_{\mathbf{L}}[\ln p(\mathbf{Z}_t, \mathbf{L}_t, x_t^i)] + \text {const} \\
&= \ln \varphi_{t}^{i}\left(x_{t}^{i}\right)_{forward} + \sum_{j=1}^{N^{mes}_t} \left( \mathbb{E}_{\mathbf{L}}[l^{j,i}] \ln \zeta_{j}(z_t^{j} \mid x_t^{i}) \right)  + \text {const.}
\end{aligned}
\label{variational_forward_x}
\end{equation} The prior of each target state can be obtained using the following equation
: \begin{equation}
\begin{aligned}
\quad \varphi_{t}^i\left(x_{t}^i\right)_{forward}&=\int \varphi_t^{i}\left(x_t^{i} \mid x_{t-1}^{i}\right) \varphi_{t-1}^{i}\left(x_{t-1}^{i}\right) d x_{t-1}^{i} \\
&= \mathcal{N}\left(x_t^{i} \mid x_{t\mid t-1}^{i},  P_{t\mid t-1}^{i}\right)
\end{aligned}
\end{equation}

With the conjugate prior we selected for target states, the forward variational distribution of each target state should also be a Gaussian distribution $q^{\star}(x_t^i)_{forward} = \mathcal{N}\left(x_t^{i} \mid x_{t\mid t}^{i},  P_{t\mid t}^{i}\right)$. The parameters of this variational distribution can be obtained by rearrange Equation (\ref{variational_forward_x}):

\begin{equation}
\begin{aligned}
&(P_{t \mid t}^{i})^{-1} = (P_{t\mid t-1}^{i})^{-1} + \sum_{j=1}^{N^{mes}_t} r_t^{j,i} (H_t^j)^T (R_t^j)^{-1} H_t^j\\
&(P_{t \mid t}^{i})^{-1} x_{t \mid t}^i = (P_{t\mid t-1}^{i})^{-1}x_{t \mid t-1}^{i} + \sum_{j=1}^{N^{mes}_t} r_t^{j,i} (H_t^j)^T (R_t^j)^{-1} z_t^j
\end{aligned}
\end{equation}

The expected values of the mixing coefficients $\pi_t^i$ can be derived by calculating the Dirichlet mean 
\begin{equation}
\begin{aligned}
&\mathbb{E}\left[\pi_t^{i}\right] = \frac{\alpha_{0}+\sum_{j=1}^{N^{mes}_t} r_t^{j,i}}{N^{tar} \alpha_{0}+N_t^{mea}}\\
&\pi_t^{N^{tar}+1} = 1-\sum_{i=1}^{N^{tar}}{\pi_t^i}
\label{variationalpara1_vpmht}
\end{aligned}
\end{equation}

\subsubsection{Backward Smoothing}
For the backward smoother, the well-known Rauch–Tung–Striebel (RTS) smoother \cite{rauch1965maximum} is applied, which assumes that the current target states are only dependent on the target states at next future time step in the smoothing process. The backward modelling of the variational distribution of target states is formulated as:

\begin{equation}
\begin{aligned}
\ln q^{\star}(x_t^i)_{backward} &= \mathbb{E}_{\mathbf{L}}[\ln p(x_t^i)] + \text {const}\\
&= \sum_{i=1}^{N^{tar}} \ln \varphi_{t}^{i}\left(x_{t}^{i}\right)_{backward}  + \text {const.}
\end{aligned}
\end{equation}where the smoothed target distribution is: \begin{equation}
\begin{aligned}
\quad \varphi_{t}^i\left(x_{t}^i\right)_{backward} &= \int \varphi_t^{i}\left(x_t^{i} \mid x_{t+1}^{i}\right) \varphi_{t+1}^{i}\left(x_{t+1}^{i}\right) d x_{t+1}^{i} \\
&= \mathcal{N}\left(x_t^{i} \mid \hat{x}_{t \mid t}^i, \hat{P}_{t \mid t}^i\right).
\end{aligned}
\end{equation} $\hat{x}_{t \mid t}^i$ and $\hat{P}_{t \mid t}^i$ are the smoothed a posterior estimate and its corresponding error covariance matrix of target $i$ at time $t$:
\begin{equation}
\begin{aligned}
\hat{x}_{t \mid t}^i &= x_{t \mid t}^i+P_{t \mid t}^i\left(F_{t+1}^{i}\right)^{T} (P_{t \mid t-1}^i)^{-1}\left[\hat{x}_{t+1 \mid t+1}^i-F_{t+1}^{i} x_{t \mid t}^i\right]\\
\hat{P}_{t \mid t}^i &=P_{t \mid t}^i+P_{t \mid t}^i\left(F_{t+1}^{i}\right)^{T} (P_{t \mid t-1}^i)^{-1} \left[\hat{P}_{t+1 \mid t+1}^i-P_{t \mid t-1}^i\right] (P_{t \mid t-1}^i)^{-1} \left(F_{t+1}^{i}\right)^{T} P_{t \mid t}^i
\end{aligned}
\end{equation}

\subsection{Stop Criterion}
To improve iteration efficiency, a good termination condition for the EM-like iteration is indispensable. In this paper, the evidence lower bound (ELOB) $\mathcal{L}$ is selected as the termination condition because of its easy availability and representativeness of the convergence. The iteration will stop when the increase of ELOB is not significant enough ($\bigtriangleup \mathcal{L} < \bigtriangleup \mathcal{L}_{terminate}$). Moreover, ELOB is also a good tool to check the software implementation of the algorithm since its value should always increase compared to the last iteration. The ELOB is given by
\begin{equation}
\begin{aligned}
\mathcal{L} &= \mathbb{E}[\ln{P(\mathbf{Z}, \mathbf{X}, \mathbf{L}, \boldsymbol{\pi})}]-\mathbb{E}[\ln q(\mathbf{L}, \boldsymbol{\pi}, \mathbf{X}) ]\\   &=\mathbb{E}[\ln p(\mathbf{Z} | \mathbf{L}, \mathbf{X})] +\mathbb{E}[\ln p(\mathbf{L} | \boldsymbol{\pi})]+\mathbb{E}[\ln p(\boldsymbol{\pi})]+\mathbb{E}[\ln p(\mathbf{X})]\\
&\quad -\mathbb{E}[\ln q(\mathbf{L})]-\mathbb{E}[\ln q(\boldsymbol{\pi})]-\mathbb{E}[\ln q(\mathbf{X})].
\end{aligned}
\end{equation} 

\subsection{VBEM Solution}
Simultaneously optimising $q^{\star}(\boldsymbol{\pi}, \mathbf{X})$ and $q^{\star}(\mathbf{L})$ can be challenging and often mathematically intractable. VBEM method tackles this issue by iteratively optimising $q^{\star}(\mathbf{L})$ using $q^{\star}(\boldsymbol{\pi}, \mathbf{X})$ from the last iteration, then updating $q^{\star}(\boldsymbol{\pi}, \mathbf{X})$ with current $q^{\star}(\mathbf{L})$ until $q(\mathbf{L}, \boldsymbol{\pi}, \mathbf{X})$ reaches convergence. The convergence criterion applied is the evidence lower bound (ELOB) $\mathcal{L}$ because of its easy availability and representativeness of the convergence. The iteration will stop when the increase of ELOB is not significant enough ($\bigtriangleup \mathcal{L} < \bigtriangleup \mathcal{L}_{terminate}$). We introduce the time batch $t_{batch}$ so that the algorithm can perform both forward filtering and backward smoothing. The VPMHT algorithm (see \textbf{Algorithm 2} for pseudo-code) at each time step can be formulated as: 

\begin{enumerate}
    \item Performing Kalman prediction to obtain the prior estimates for all time steps in the batch, $t = t, t+1, ..., t+t_{batch}$: 
    \begin{equation}
    \begin{aligned}
    &x_{t\mid t-1}^{i} = F_t^i \hat{x}_{t-1\mid t-1}^{i} \\
    &P_{t\mid t-1}^{i} = F_t^i \hat{P}_{t-1\mid t-1}^{i} (F_t^i)^{T} + Q_t^i \\
    \end{aligned}    
    \end{equation}where ${x}_{t|t-1}^i$ and ${P}_{t|t-1}^i$ are the prior state estimate and its corresponding error covariance matrix for $x_t^i$. ${x}_{t|t}^i$ and ${P}_{t|t}^i$ are the a posterior state estimate and its corresponding error covariance matrix.
    \item E-step: Optimising $q^{\star}(\mathbf{L})$ using $q^{\star}(\boldsymbol{\pi}, \mathbf{X})$ from last iteration for all time steps in the batch, $t = t, t+1, ..., t+t_{batch}$.
    
    Here, we introduce the responsibility parameter $r_t^{j,i}$ that satisfies $\mathbb{E}\left[l^{j,i}\right] = r^{j,i}$ to give a well representation of $q^{\star}(\mathbf{L})$:
    \begin{equation}
    \begin{aligned}
    &r_t^{j,i} = \frac{\rho_t^{j,i}}{\sum_{i=1}^{N^{tar}+1} \rho_t^{j,i}} \\
    \end{aligned}
    \end{equation}
    with
    \begin{equation}
    \begin{aligned}
    &\ln \rho_t^{j,i}=\left\{\begin{array}{ll}
    \mathbb{E}_{\boldsymbol{\pi}}[\ln{ \pi_t^i}] - \frac{D_t^j}{2}\ln{2\pi} -  \frac{D_t^j}{2}\ln{\operatorname{det}(R_t^j)} \\
    -\frac{1}{2} \mathbb{E}_{\mathbf{X}}[(H_t^j x_t^i - z_t^j)^{T}(R_t^j)^{-1}(H_t^j x_t^i - z_t^j)] & \text { if } i \in \{1, \ldots, N^{tar}\} \\ 
    \frac{\pi_t^{N^{tar}+1}}{B-A} & \text { if } i =N^{tar}+1
    \end{array}\right.\\
    &\mathbb{E}_{\boldsymbol{\pi}}\left[\ln \pi_t^{i}\right] = \psi\left(\alpha^i\right)-\psi(\widehat{\alpha})\\
    &\mathbb{E}_{\mathbf{X}}[(H_t^j x_t^i - z_t^j)^{T}(R_t^j)^{-1}(H_t^j x_t^i - z_t^j)]& \\
    &\quad \quad \quad = \left(H_t^j x_{t \mid t}^i-z_t^j\right)^{\mathrm{T}}(R_t^j)^{-1}\left(H_t^j x_{t \mid t}^i-z_t^j\right)+ \operatorname{Tr}((H_t^j)^T (R_t^j)^{-1} H_t^j P_{t \mid t}^{i})
    \end{aligned}
    \end{equation}where $D_t^j$ is the dimension of measurement ${z}_t^{j}$. $\operatorname{det}(\cdot))$ is the determinant operator and $\operatorname{Tr}(\cdot))$ is the trace operator. $\psi(\cdot)$ is the digamma function $\psi(\alpha) = \frac{d}{d \alpha} \ln \Gamma(\alpha)$, $\Gamma(\alpha)=(\alpha-1) !$, and $\widehat{\alpha_t}=\sum\alpha_t^{i}$. 
    
    \item M-step: Updating $q^{\star}(\boldsymbol{\pi}, \mathbf{X})$ with $q^{\star}(\mathbf{L})$ from E-step.
    \begin{itemize}
        \item Forward M-step \\
        Computing variational parameters $\pi_t^i$, $x_{t \mid t}^i$, $P_{t \mid t}^i$ using $r_t^{j,i}$ from E-step for all times steps in the batch $t = t, t+1, ..., t+t_{batch}$\\
        \begin{equation}
        \begin{aligned}
        &(P_{t \mid t}^{i})^{-1} = (P_{t\mid t-1}^{i})^{-1} + \sum_{j=1}^{N^{mes}_t} r_t^{j,i} (H_t^j)^T (R_t^j)^{-1} H_t^j\\
        &(P_{t \mid t}^{i})^{-1} x_{t \mid t}^i = (P_{t\mid t-1}^{i})^{-1}x_{t \mid t-1}^{i} + \sum_{j=1}^{N^{mes}_t} r_t^{j,i} (H_t^j)^T (R_t^j)^{-1} z_t^j\\
        &\pi_t^{i} =\left\{\begin{array}{ll} \frac{\alpha_{0}+\sum_{j=1}^{N^{mes}_t} r_t^{j,i}}{N^{tar} \alpha_{0}+N_t^{mea}} & \text { if } i \in \{1, \ldots, N^{tar}\} \\
        1-\sum_{i=1}^{N^{tar}}{\pi_t^i} & \text { if } i =N^{tar}+1
        \end{array}\right.
        \end{aligned}
        \end{equation}
        \item Backward M-step \\
        Computing smoothed variational parameters ${\pi}_t^i$, $\hat{x}_{t \mid t}^i$, and $\hat{P}_{t \mid t}^i$ by the backward recursion for $t = t+t_{batch}-1, t+t_{batch}-2, ..., t$ \\
        \begin{equation}
        \begin{aligned}
        \hat{x}_{t \mid t}^i &= x_{t \mid t}^i+P_{t \mid t}^i\left(F_{t+1}^{i}\right)^{T} (P_{t \mid t-1}^i)^{-1}\left[\hat{x}_{t+1 \mid t+1}^i-F_{t+1}^{i} x_{t \mid t}^i\right]\\
        \hat{P}_{t \mid t}^i &=P_{t \mid t}^i+P_{t \mid t}^i\left(F_{t+1}^{i}\right)^{T} (P_{t \mid t-1}^i)^{-1} \left[\hat{P}_{t+1 \mid t+1}^i-P_{t \mid t-1}^i\right] (P_{t \mid t-1}^i)^{-1} \left(F_{t+1}^{i}\right)^{T} P_{t \mid t}^i
        \end{aligned}
        \end{equation}
    \end{itemize}
    \item Returning to step 2 for the next VBEM iteration, unless the ELOB $\mathcal{L}$ reaches the stopping criterion $\bigtriangleup \mathcal{L} < \bigtriangleup \mathcal{L}_{terminate}$ at the current time step $t$:
    \begin{equation}
    \begin{aligned}
    \mathcal{L} &= \mathbb{E}[\ln{P(\mathbf{Z}, \mathbf{X}, \mathbf{L}, \boldsymbol{\pi})}]-\mathbb{E}[\ln q(\mathbf{L}, \boldsymbol{\pi}, \mathbf{X}) ]\\   &=\mathbb{E}[\ln p(\mathbf{Z} | \mathbf{L}, \mathbf{X})] +\mathbb{E}[\ln p(\mathbf{L} | \boldsymbol{\pi})]+\mathbb{E}[\ln p(\boldsymbol{\pi})]+\mathbb{E}[\ln p(\mathbf{X})]\\
    &\quad -\mathbb{E}[\ln q(\mathbf{L})]-\mathbb{E}[\ln q(\boldsymbol{\pi})]-\mathbb{E}[\ln q(\mathbf{X})].
    \end{aligned}
    \end{equation}
\end{enumerate}

\begin{algorithm*}
\caption{Variational PMHT algorithm at scan $t$}
\begin{algorithmic}[1]
\renewcommand{\algorithmicrequire}{\textbf{Input:}}
\renewcommand{\algorithmicensure}{\textbf{Output:}}
\REQUIRE The bounds of the clutter distribution $(A_t,B_t)$, number of targets $N^{tar}$, the smoothed previous estimation $\left\{\hat{x}_{t-1 \mid t-1}^{i}, \hat{P}_{t-1 \mid t-1}^{i}\right\},$ received measurements $\mathbf{Z}$, and termination condition $\bigtriangleup \mathcal{L}_{terminate}$
\ENSURE  The smoothed current estimation $\{\hat{x}_{t \mid t}^i, \hat{P}_{t \mid t}^i\}$
\WHILE{$\bigtriangleup \mathcal{L} > \bigtriangleup \mathcal{L}_{terminate}$}
\FOR{$t = t, \ldots, T$}
\STATE Kalman Predict \\
\quad
\begin{math}
\begin{aligned}
&x_{t\mid t-1}^{i} = F_t^i \hat{x}_{t-1\mid t-1}^{i} \\
&P_{t\mid t-1}^{i} = F_t^i \hat{P}_{t-1\mid t-1}^{i} (F_t^i)^{T} + Q_t^i \\
\end{aligned}
\end{math}
\STATE E-step \\
Updating responsibilities $r_t^{j,i}$ with variational parameters from the last iteration\\
\quad
\begin{math}
\begin{aligned}
r_t^{j,i} = \frac{\rho_t^{j,i}}{\sum_{i=1}^{N^{tar}+1} \rho_t^{j,i}}
\end{aligned}
\end{math}\\
\quad where
\\ \quad
\begin{math}
\begin{aligned}
&\ln \rho_t^{j,i}=\left\{\begin{array}{ll}
\mathbb{E}_{\boldsymbol{\pi}}[\ln{ \pi_t^i}] - \frac{D_t^j}{2}\ln{2\pi} -  \frac{D_t^j}{2}\ln{\operatorname{det}(R_t^j)} \\
-\frac{1}{2} \mathbb{E}_{\mathbf{X}}[(H_t^j x_t^i - z_t^j)^{T}(R_t^j)^{-1}(H_t^j x_t^i - z_t^j)] & \text { if } i \in \{1, \ldots, N^{tar}\} \\ 
\frac{\pi_t^{N^{tar}+1}}{B-A} & \text { if } i =N^{tar+1}
\end{array}\right.\\
&\mathbb{E}_{\boldsymbol{\pi}}\left[\ln \pi^{i}\right] = \psi\left(\alpha^i\right)-\psi(\widehat{\alpha})\\
&\mathbb{E}_{\mathbf{X}}[(H_t^j x_t^i - z_t^j)^{T}(R_t^j)^{-1}(H_t^j x_t^i - z_t^j)]& \\
&\quad \quad \quad = \left(H_t^j x_{t \mid t}^i-z_t^j\right)^{\mathrm{T}}(R_t^j)^{-1}\left(H_t^j x_{t \mid t}^i-z_t^j\right)+ \operatorname{Tr}((H_t^j)^T (R_t^j)^{-1} H_t^j P_{t \mid t}^{i})
\end{aligned}
\end{math}
\STATE Forward M-step \\
Computing variational parameters $\pi_t^i$, $x_{t \mid t}^i$, $P_{t \mid t}^i$ using $r_t^{j,i}$ from E-step \\
\quad
\begin{math}
\begin{aligned}
&(P_{t \mid t}^{i})^{-1} = (P_{t\mid t-1}^{i})^{-1} + \sum_{j}^{N^{mes}_t} r_t^{j,i} (H_t^j)^T (R_t^j)^{-1} H_t^j\\
&(P_{t \mid t}^{i})^{-1} x_{t \mid t}^i = (P_{t\mid t-1}^{i})^{-1}x_{t \mid t-1}^{i} + \sum_{j}^{N^{mes}_t} r_t^{j,i} (H_t^j)^T (R_t^j)^{-1} z_t^j\\
&\pi_t^{i} =\left\{\begin{array}{ll} \frac{\alpha_{0}+\sum_{j}^{N^{mes}_t} r_t^{j,i}}{N^{tar} \alpha_{0}+N_t^{mea}} & \text { if } i \in \{1, \ldots, N^{tar}\} \\
1-\sum_{i=1}^{N^{tar}}{\pi_t^i} & \text { if } i =N^{tar+1}
\end{array}\right.
\end{aligned}
\end{math}\qquad 
\ENDFOR
\FOR{$t = T, \ldots, t$}
\STATE Backward M-step \\
Computing smoothed state estimates $\hat{x}_{t \mid t}^i$ and covariances $\hat{P}_{t \mid t}^i$ \\
\quad
\begin{math}
\begin{aligned}
\hat{x}_{t \mid t}^i &= x_{t \mid t}^i+P_{t \mid t}^i\left(F_{t+1}^{i}\right)^{T} (P_{t \mid t-1}^i)^{-1}\left[\hat{x}_{t+1 \mid t+1}^i-F_{t+1}^{i} x_{t \mid t}^i\right]\\
\hat{P}_{t \mid t}^i &=P_{t \mid t}^i+P_{t \mid t}^i\left(F_{t+1}^{i}\right)^{T} (P_{t \mid t-1}^i)^{-1} \left[\hat{P}_{t+1 \mid t+1}^i-P_{t \mid t-1}^i\right] (P_{t \mid t-1}^i)^{-1} \left(F_{t+1}^{i}\right)^{T} P_{t \mid t}^i
\end{aligned}
\end{math}
\ENDFOR
\STATE Computing ELOB $\mathcal{L}$ as the stop criterion\\ \quad 
\begin{math} \begin{aligned} 
\mathcal{L} = \mathbb{E}[\ln p(\mathbf{X}, \mathbf{L}, \boldsymbol{\pi}, \boldsymbol{\mu}, \boldsymbol{\Lambda})]-\mathbb{E}[\ln q(\mathbf{L}, \boldsymbol{\pi}, \boldsymbol{\mu}, \boldsymbol{\Lambda})] 
\label{ELOB_vgmm1}
\end{aligned}
\end{math}
\ENDWHILE
\end{algorithmic}
\end{algorithm*}

\newpage
\section{Performance Analysis}
\subsection{Some remarks}
Conventional MTT algorithms, such as probabilistic data association filter (PDAF), carry out the MTT tasks through a two-step approach.  Firstly, the measurement-to-target association likelihood is calculated. The target state estimates are then computed with a fusion algorithm using the association likelihood and associated measurements. Apart from the iterative manner of the proposed VPMHT, the E-step and M-step share many similarities to the association step and the fusion step of PDAF-like MTT algorithms. 

The E-step calculates the responsibility $r_t^{j, i}$ with the combination of a scaled Mahalanobis distance and a factor of prior association probability. The calculation of responsibilities is very similar to the calculation of association likelihood of PDAF, which is done using a scaled Mahalanobis distance. 

As for the M-step, the equations are almost identical to the ones in Bayesian optimal Kalman filter with multiple measurements \cite{kay1993fundamentals}. If the responsibilities in the M-step correctly represent the measurement-to-target association, the M-step of VPMHT is the same as Bayesian optimal Kalman filter with multiple measurements. This fusion scheme of the VPMHT is more advanced than the ones in PMHT from the aspect of utilising the measurement uncertainties and enabling the implementation of different measurement models. The following remark can be offered from the observation of M-step of VPMHT. 

\theoremstyle{remark}
\begin{remark}
The M-step of VPMHT is identical to the Bayesian optimal Kalman filter with multiple measurements if the measurement-to-target association is correct. This implies that the fusion in VPMHT would be more advanced than the weighted sum of measurements and posterior estimates applied in PMHT and PDAF. Therefore, it can be concluded that in scenarios with straight-forward associations, the proposed VPMHT would outperform the PMHT and PDAF thanks to its fusion advancement.
\end{remark}

\subsection{Simulation Studies}
In this section, the performance of the proposed VPMHT algorithm is evaluated with extensive numerical simulations. The proposed algorithm is compared with its baseline method PMHT to evaluate its advancements and limitations. PDAF \cite{musicki1994integrated} is also considered in simulation studies for its common usage in evaluating PMHT-like algorithms \cite{willett1998variety, Willett2002pmht}. Since this paper mainly focuses on the track-loss handling of MTT algorithms, a sophisticated track management system may not fully examine the track-loss handling potential of the considered algorithms. We use a simple termination condition to identify track-losses: a target is considered lost if it is not well detected for $3$ consecutive frames. The tracking accuracy is quantified by the optimal sub-pattern assignment (OSPA) distance metric \cite{schuhmacher2008consistent}.

\subsubsection{Simulation Setup}
\begin{figure}[H]
      \centering
      \includegraphics[scale=0.6]{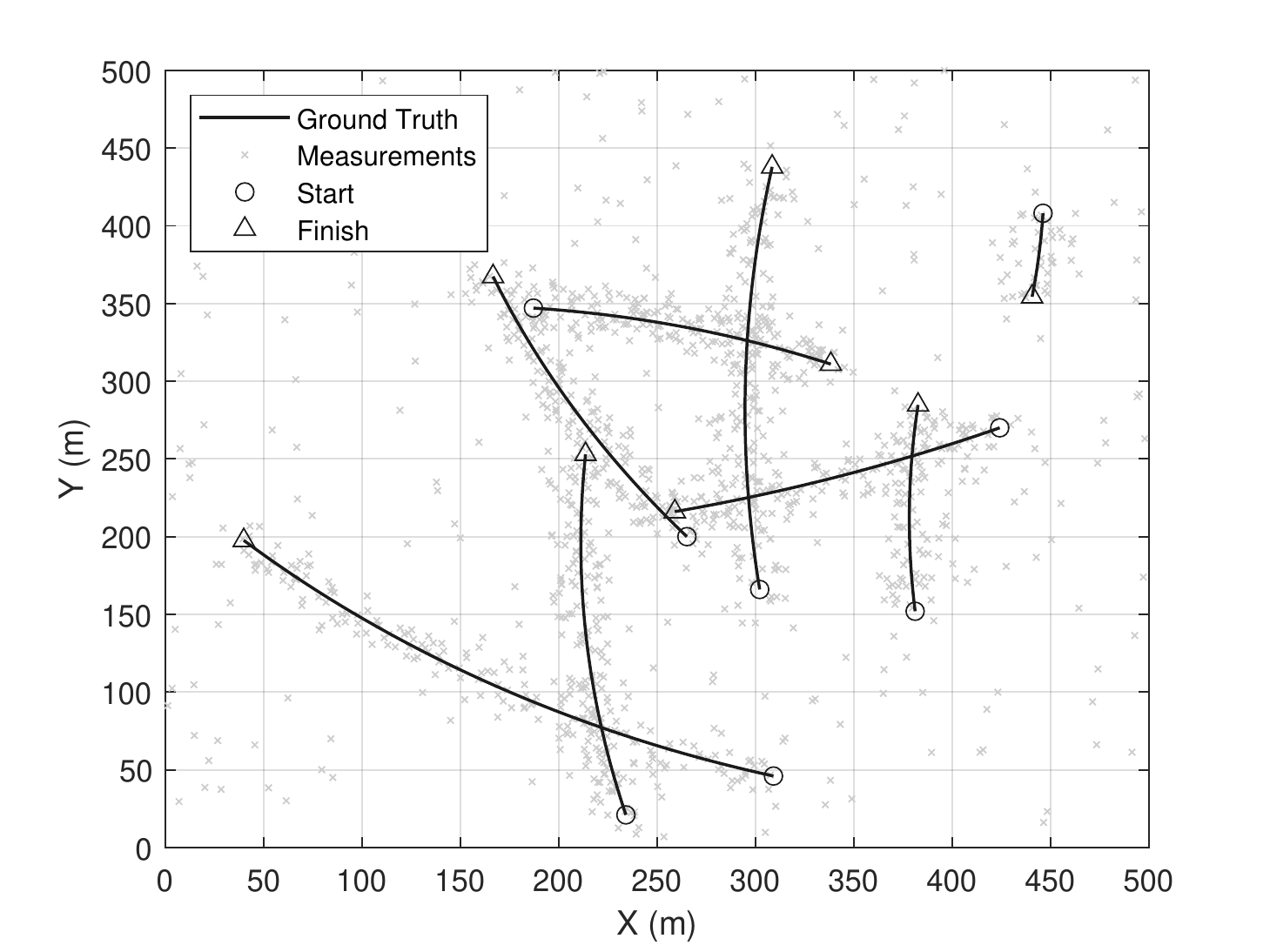}
      \caption{An example of the simulation scenario}
\label{Sim_example_vpmht}
\end{figure}

We consider a complex simulation scenario with $8$ targets randomly moves in a $500\times500 \, \text{m}^2$ square region for $40$ time steps. A single sensor monitors the whole region with measurement noise and background clutters. The clutters are assumed to be uniformly distributed in the surveillance region with its number $N_{clu}$ following Poisson. Each target can generate an arbitrary number of measurements which follows Poisson with expectation $10$. The state vector of each target $x_k^i$ is a 4-D vector with $x_k^i(1), x_k^i(3)$ position and $x_k^i(2), x_k^i(4)$ velocity components. We assume that all targets share the same constant velocity model for transition function:
\begin{equation}
x_t^{i}=F_t x_{t-1}^{i}+w_t
\label{state_measure_model_vpmht}
\end{equation} with \begin{equation}
F_t=\left[\begin{array}{cccc}
1 & 0 & 1 & 0 \\
0 & 0 & 1 & 0 \\
0 & 1 & 0 & 1 \\
0 & 0 & 0 & 1
\end{array}\right], \quad w_t \sim \mathcal{N}(0, Q_t), \quad
Q_t = \left[\begin{array}{cccc}
10 & 0 & 0 & 0 \\
0 & 1 & 0 & 0 \\
0 & 0 & 10 & 0 \\
0 & 0 & 0 & 1
\end{array}\right].\end{equation} For each target, the corresponding measurements are generated with a detection probability $P_d$ using the same measurement model:
\begin{equation}
\left. z_t^{j} \right|_{j=i}=H_t x_t^i + v_t
\end{equation}
with \begin{equation}
H_t=\left[\begin{array}{cccc}
1 & 0 & 0 & 0 \\
0 & 0 & 1 & 0 
\end{array}\right], \quad v_t \sim \mathcal{N}(0, R_t), \quad
Q_t = \delta_z^2 \left[\begin{array}{cccc}
1 & 0 \\
0 & 1 
\end{array}\right].\end{equation}where $\delta_z^2$ is the noise level. An example of simulation scenarios is presented in Figure \ref{Sim_example_vpmht}. 

In this paper, the effects of sensor noise level and clutter intensity are studied. The cases studying the effect of noise vary the noise level $\delta_z^2 = \{ 0,1,2,3,4,5\}$ and take a constant average clutters returns $N_{clu} = 10$. As for the cases with different number of clutters, we take the value of $N_{clu} = \{0,5,10,15,20,25,30\}$ with a noise level $\delta_z^2 = 3$. Scenarios with and without track-loss are both take into account. For the cases with track-loss, one random target would cease to exist for every $10$ frames. For fair comparisons, all the design parameters are set the same with time batch $T=3$, detection probability $P_d = 0.9$, and stop criterion $\bigtriangleup \mathcal{L}_{terminate} = 10^{-8}$. $200$ Monte Carlo runs are performed for each condition. 

\subsubsection{Result and Discussion}
\begin{figure}[H]
    \centering
    \subfloat[$\delta_z = 0$]{{\includegraphics[scale=.45]{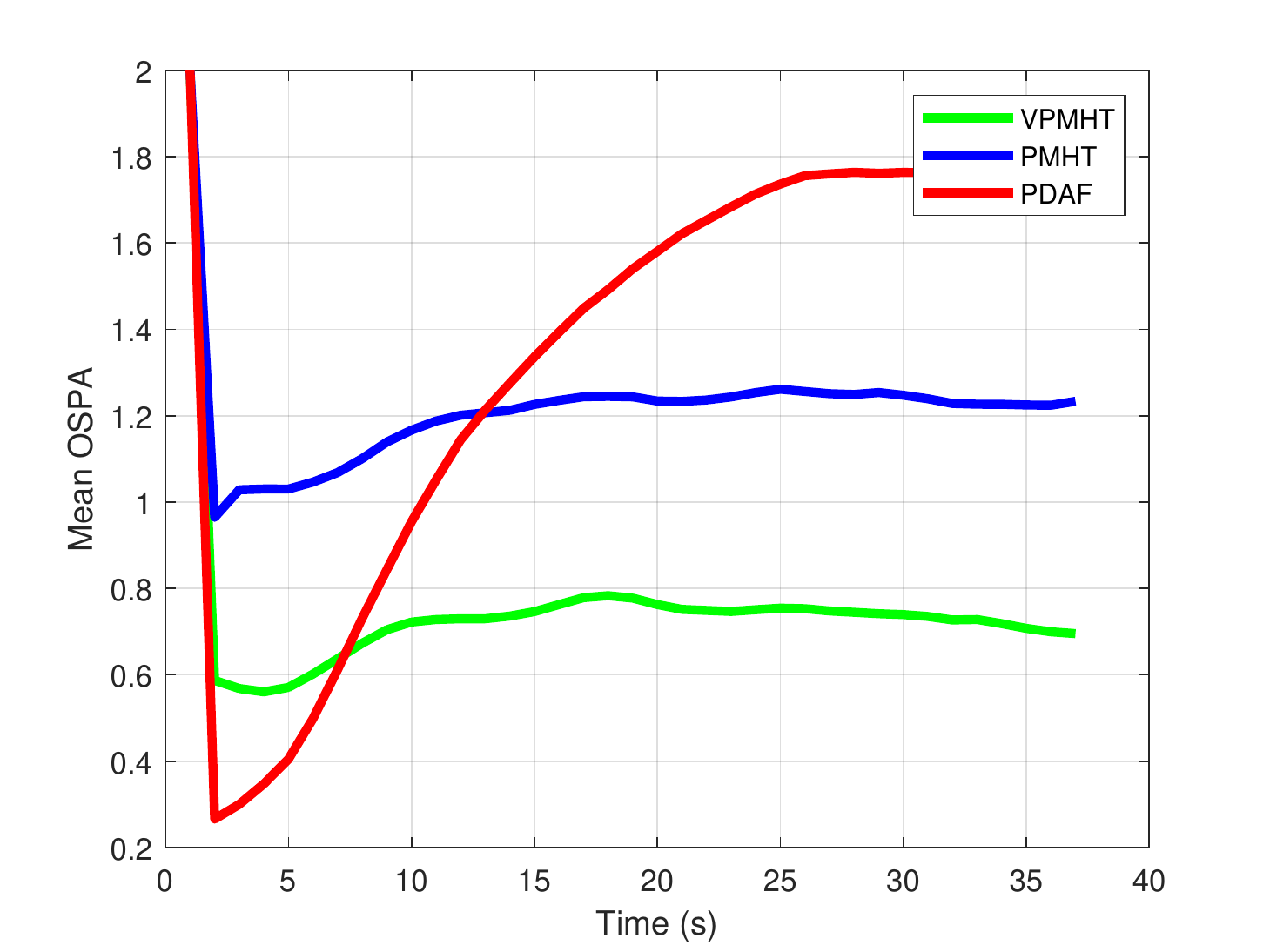}}}
    \subfloat[$\delta_z = 1$]{{\includegraphics[scale=.45]{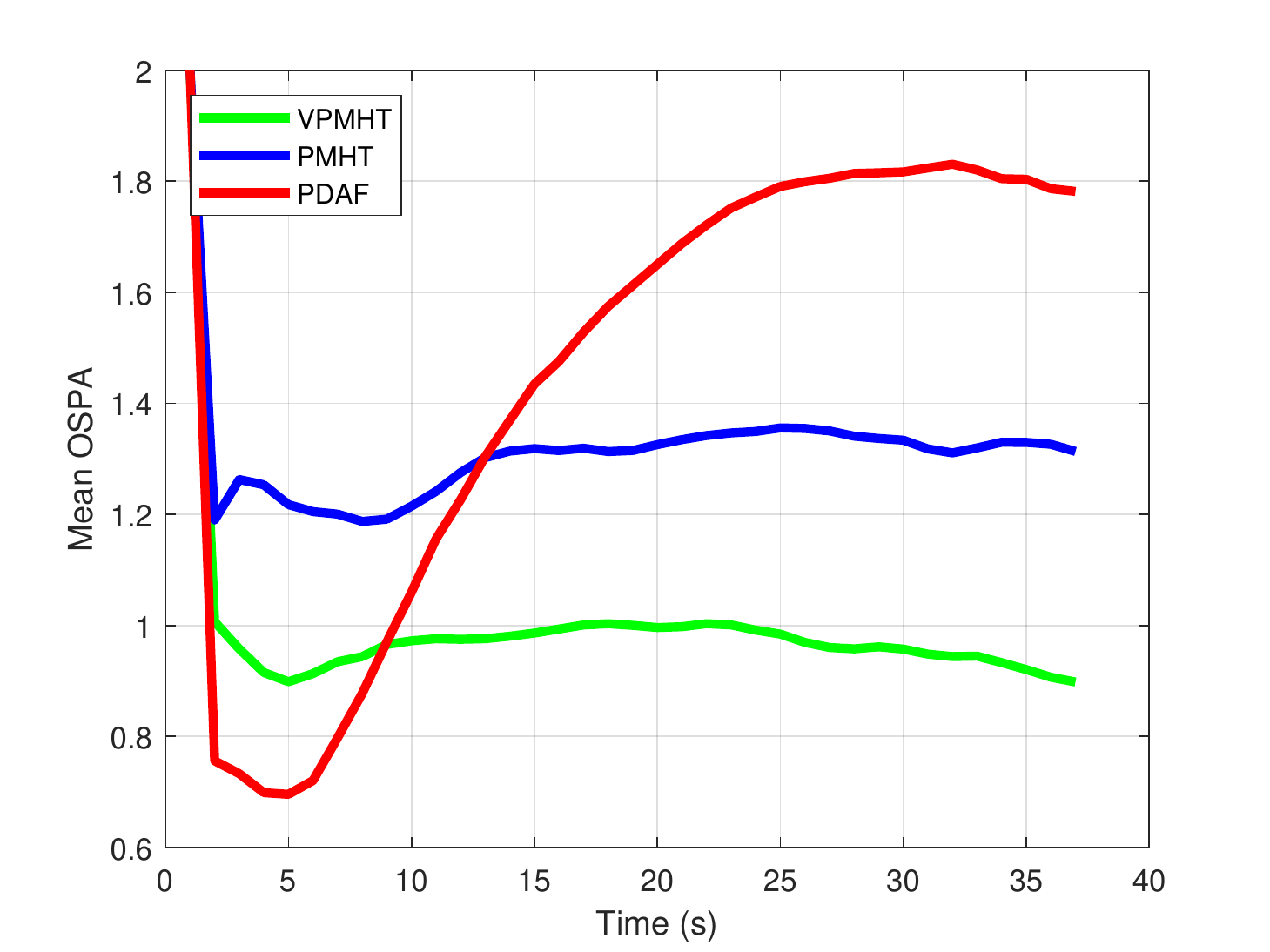}}} \\
    \subfloat[$\delta_z = 2$]{{\includegraphics[scale=.45]{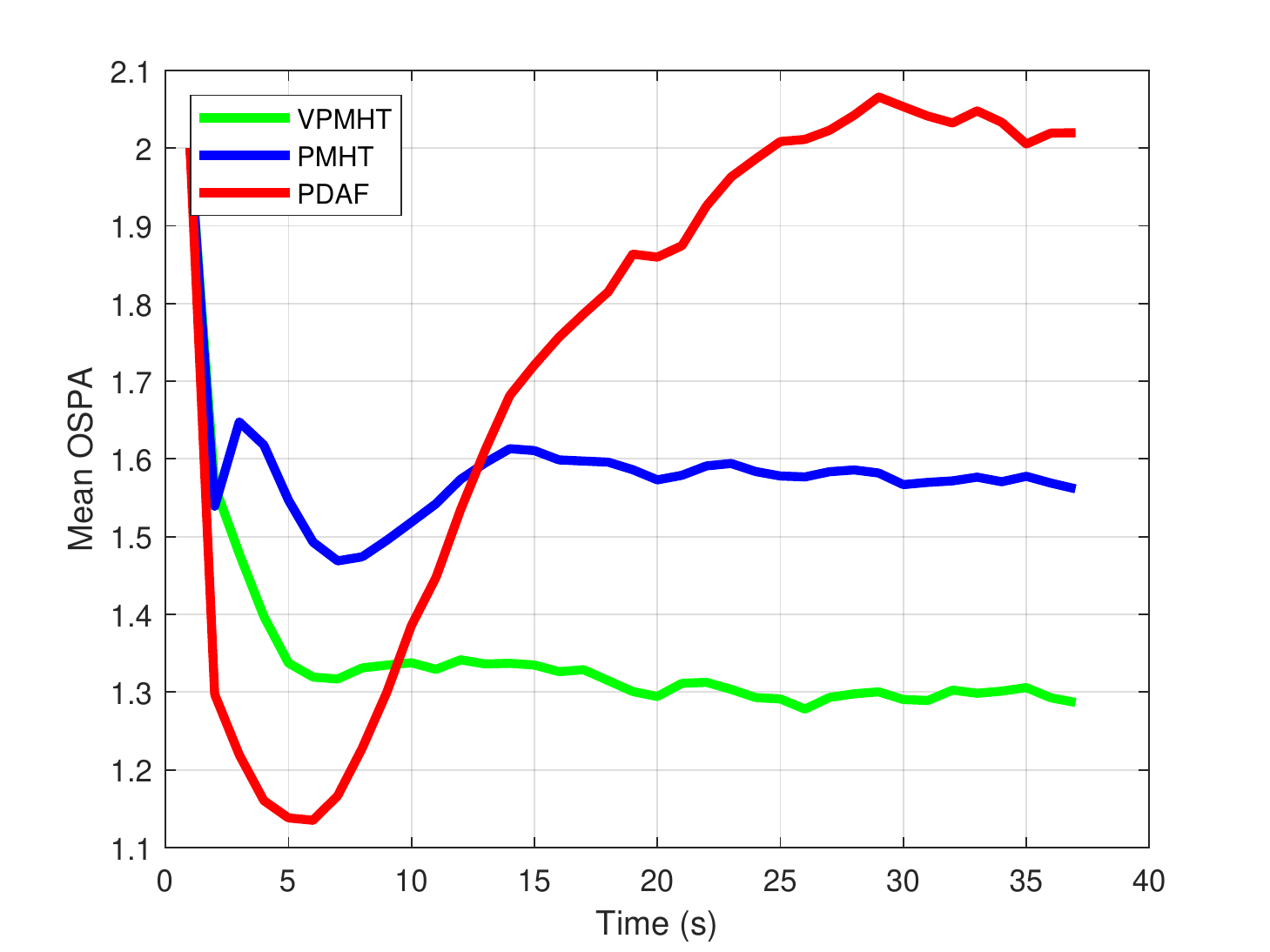}}} 
    \subfloat[$\delta_z = 3$]{{\includegraphics[scale=.45]{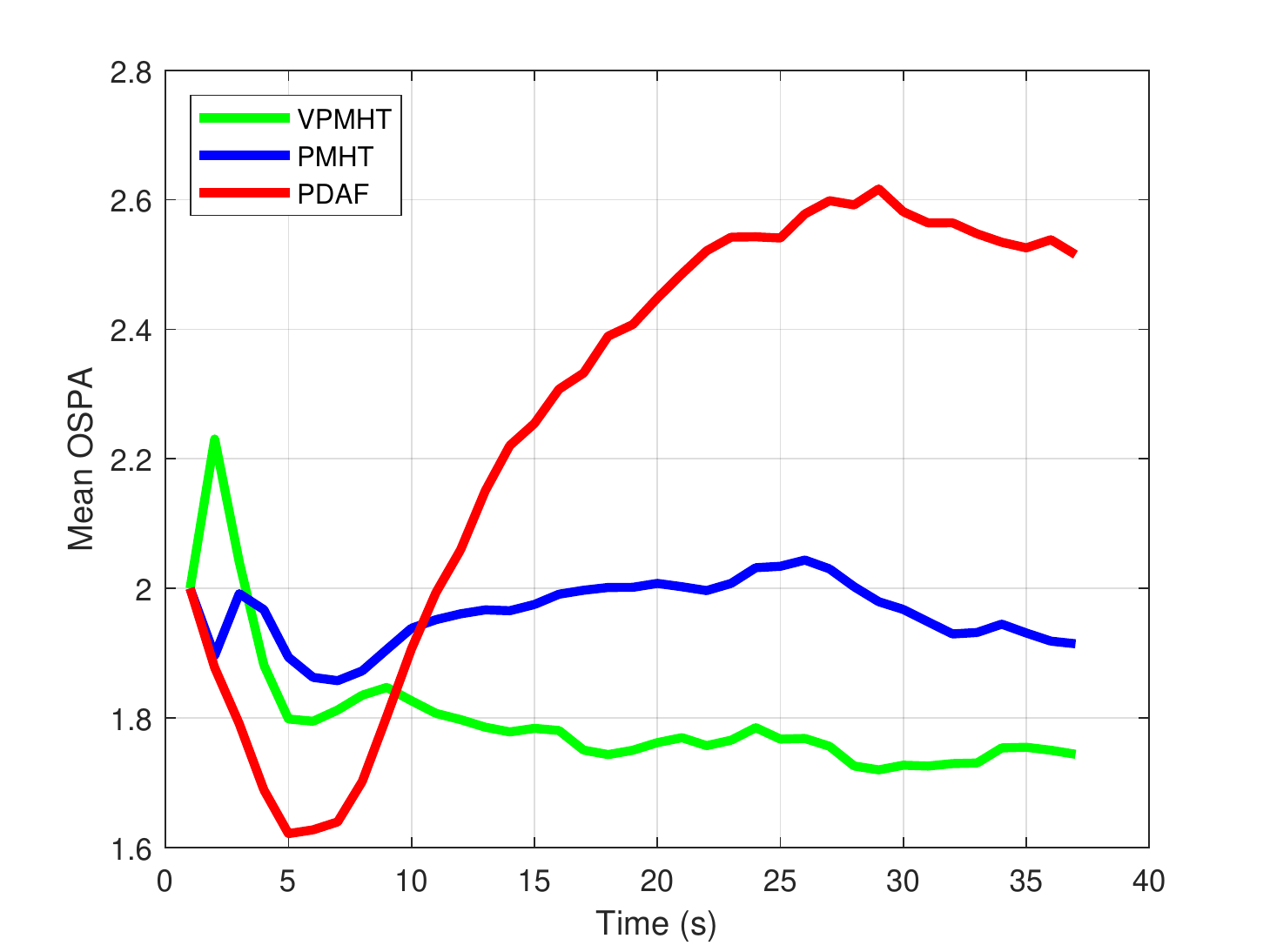}}} \\
    \subfloat[$\delta_z = 4$]{{\includegraphics[scale=.45]{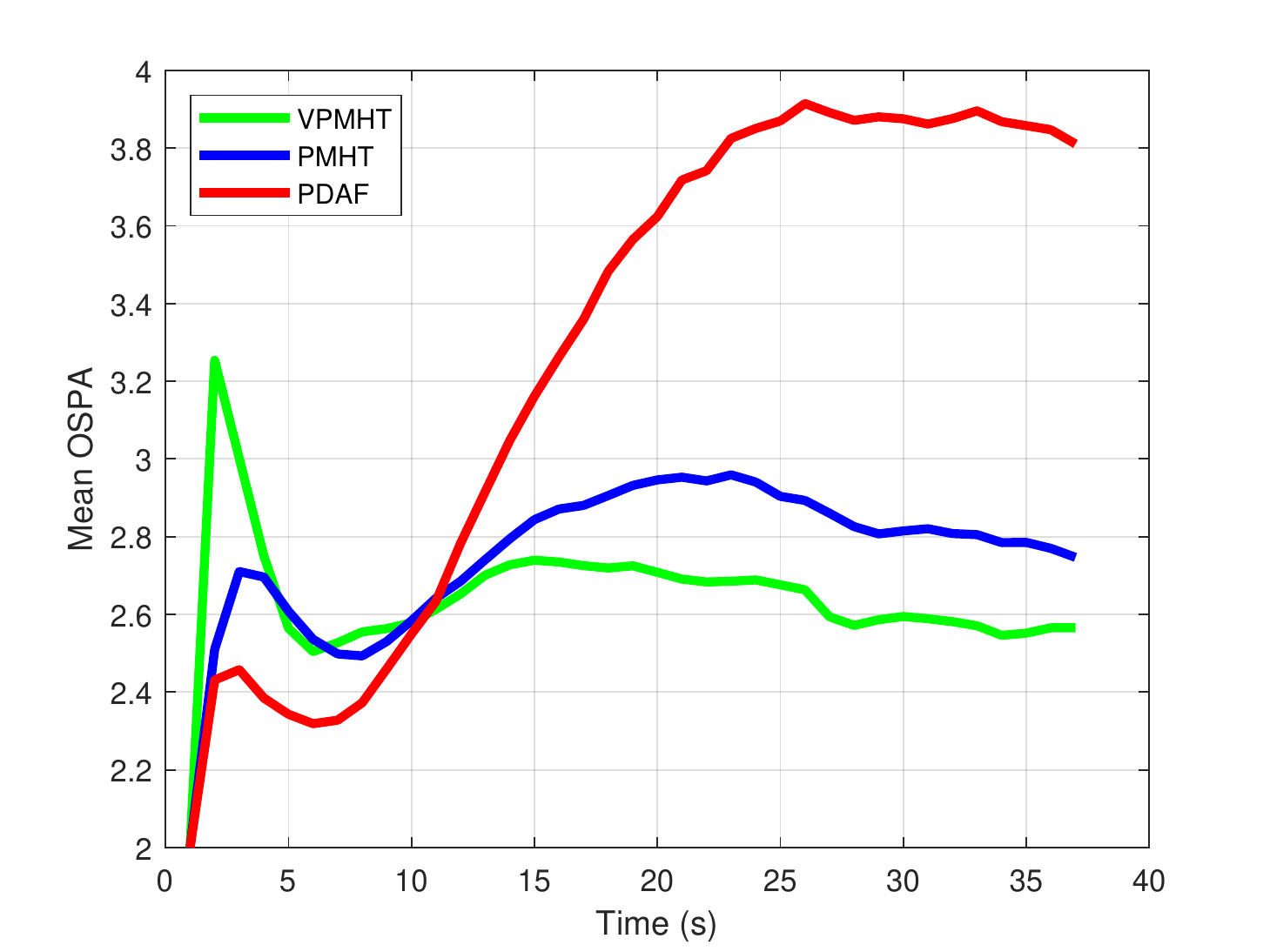}}} 
    \subfloat[Different noise level]{{\includegraphics[scale=.45]{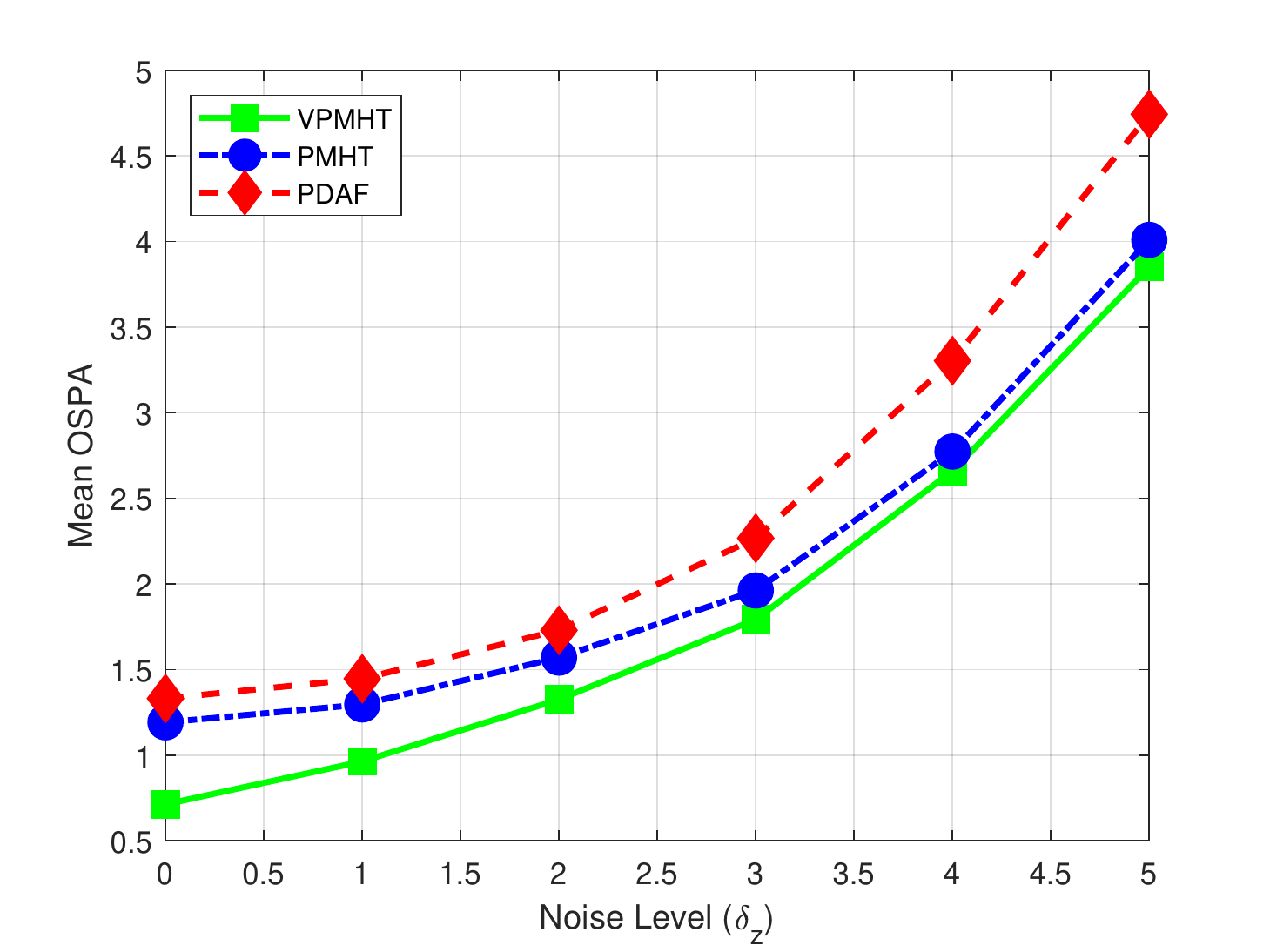}}}
    \caption{Mean OSPA w.r.t different noise level without track-loss}%
    \label{substantial_noise_vpmht}%
\end{figure}

\begin{figure}[H]
    \centering
    \subfloat[$\delta_z = 0$]{{\includegraphics[scale=.45]{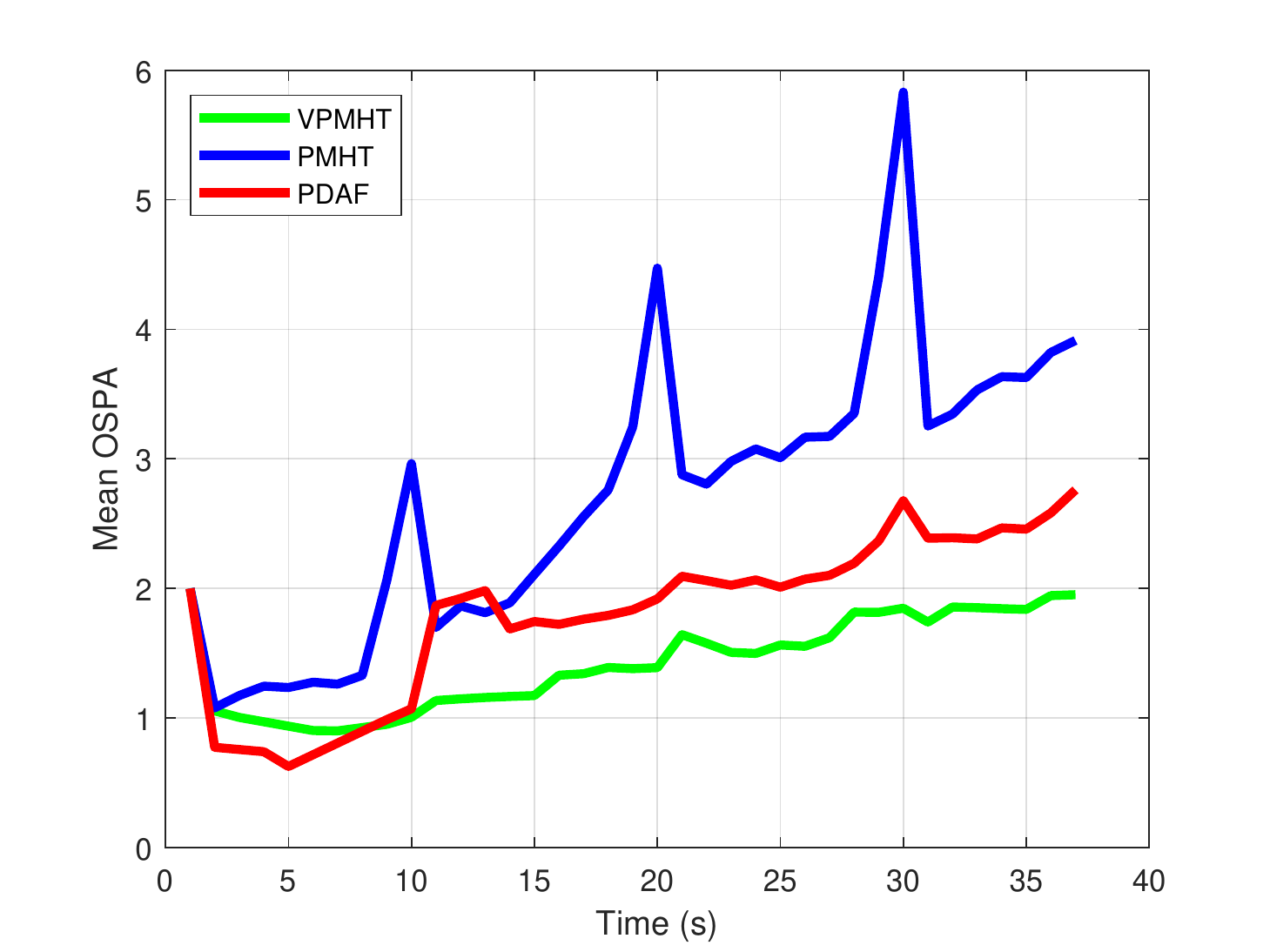}}}
    \subfloat[$\delta_z = 1$]{{\includegraphics[scale=.45]{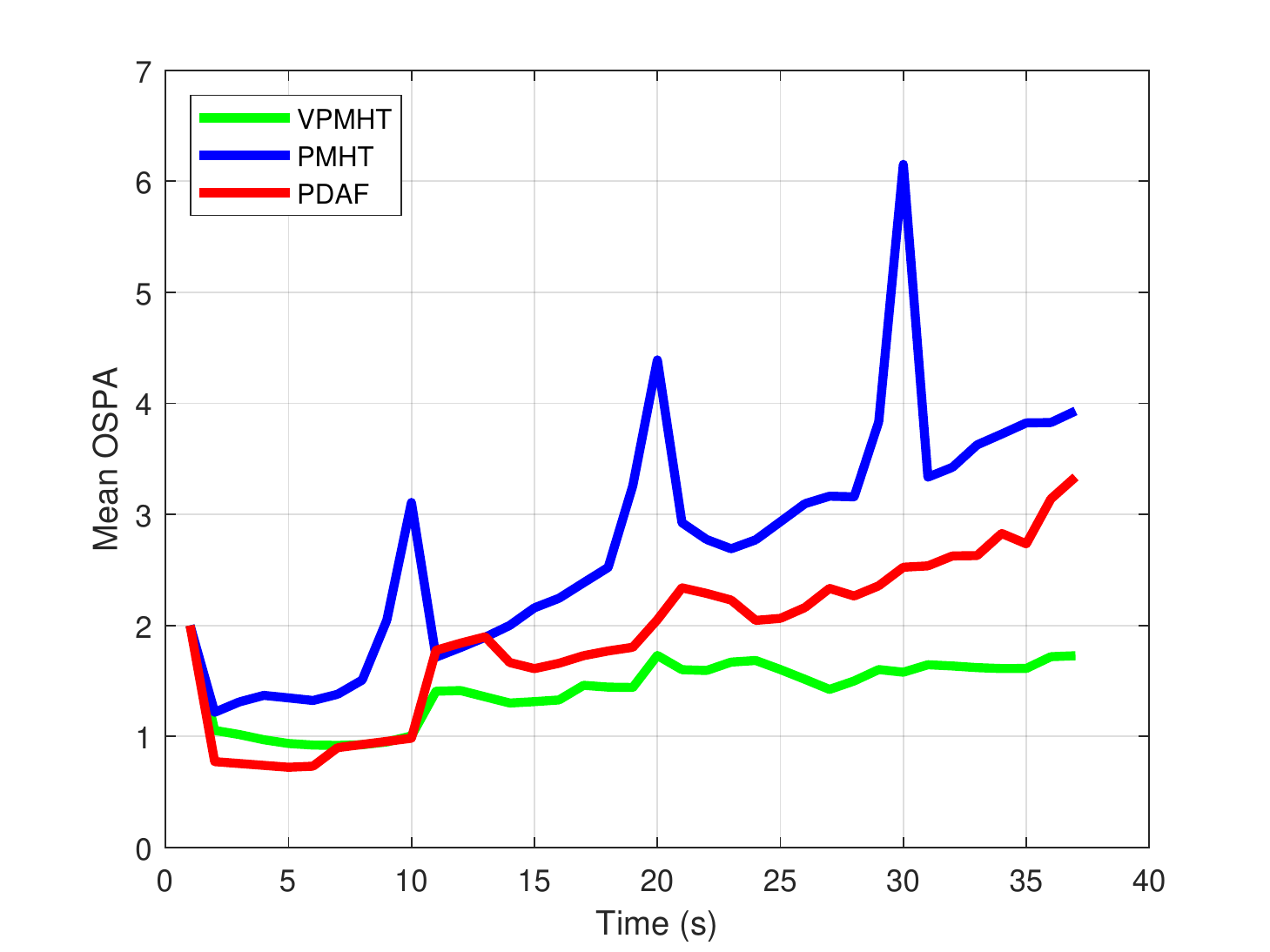}}} \\
    \subfloat[$\delta_z = 2$]{{\includegraphics[scale=.45]{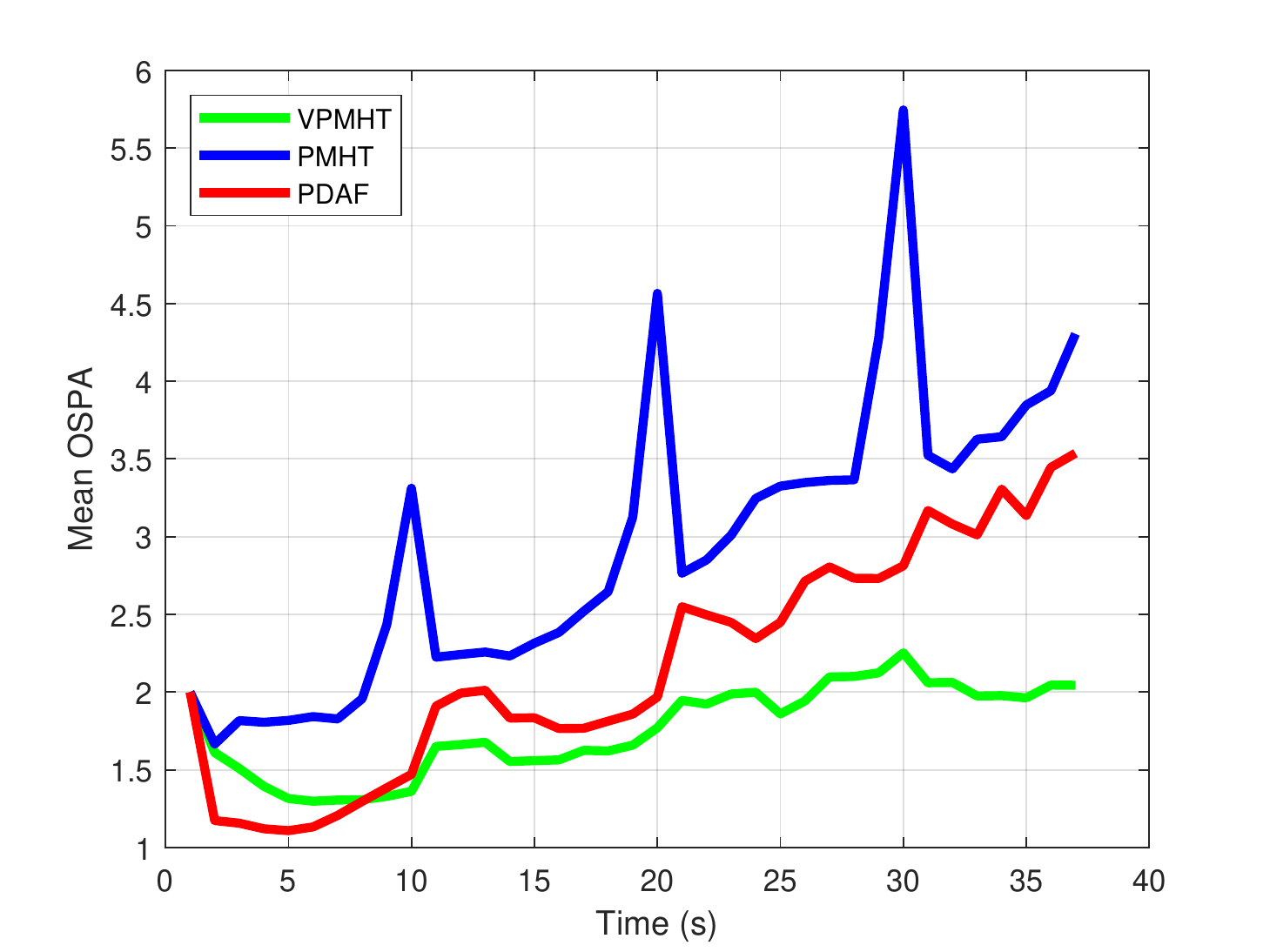}}} 
    \subfloat[$\delta_z = 3$]{{\includegraphics[scale=.45]{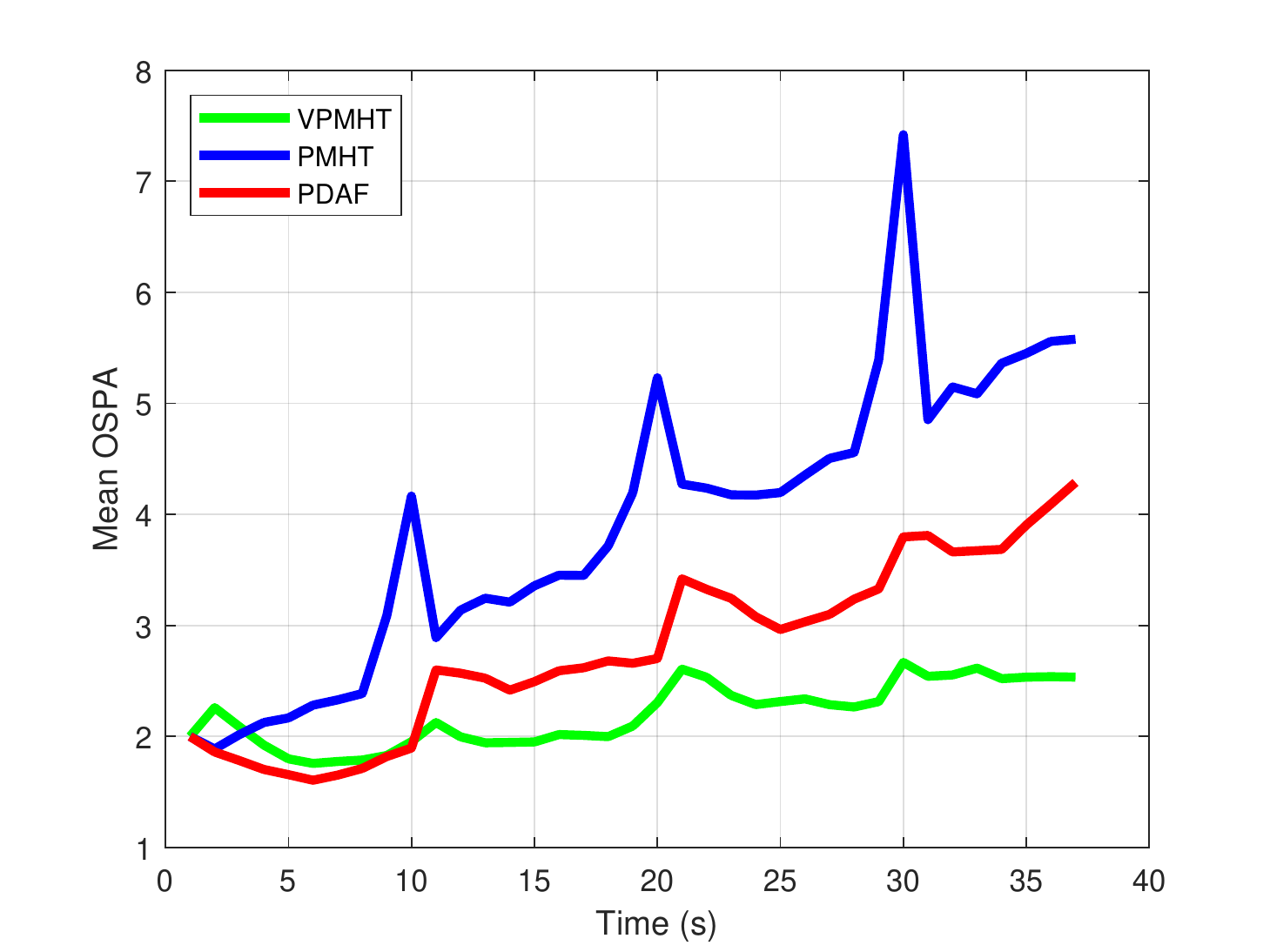}}} \\
    \subfloat[$\delta_z = 4$]{{\includegraphics[scale=.45]{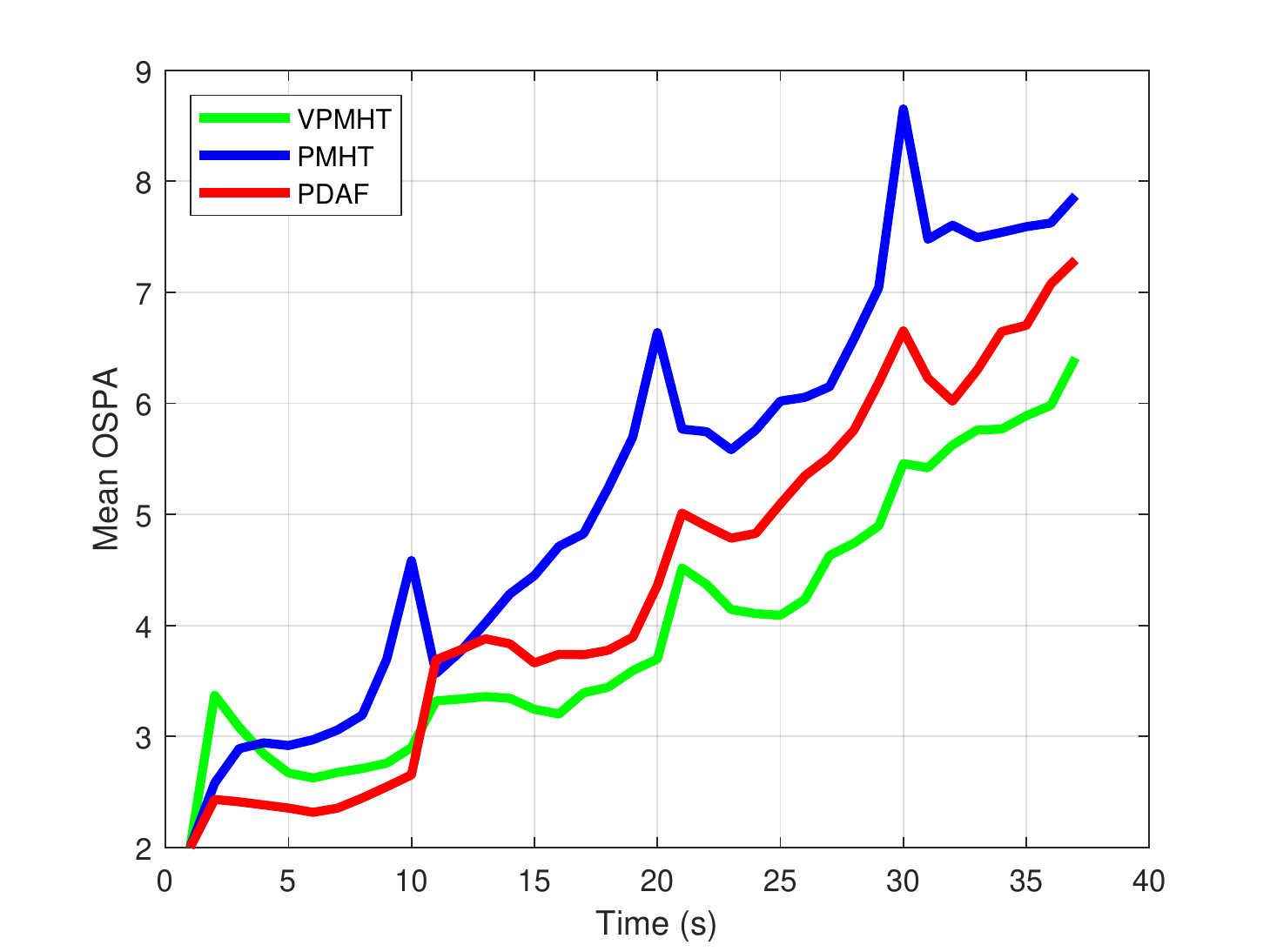}}} 
    \subfloat[Different noise level]{{\includegraphics[scale=.45]{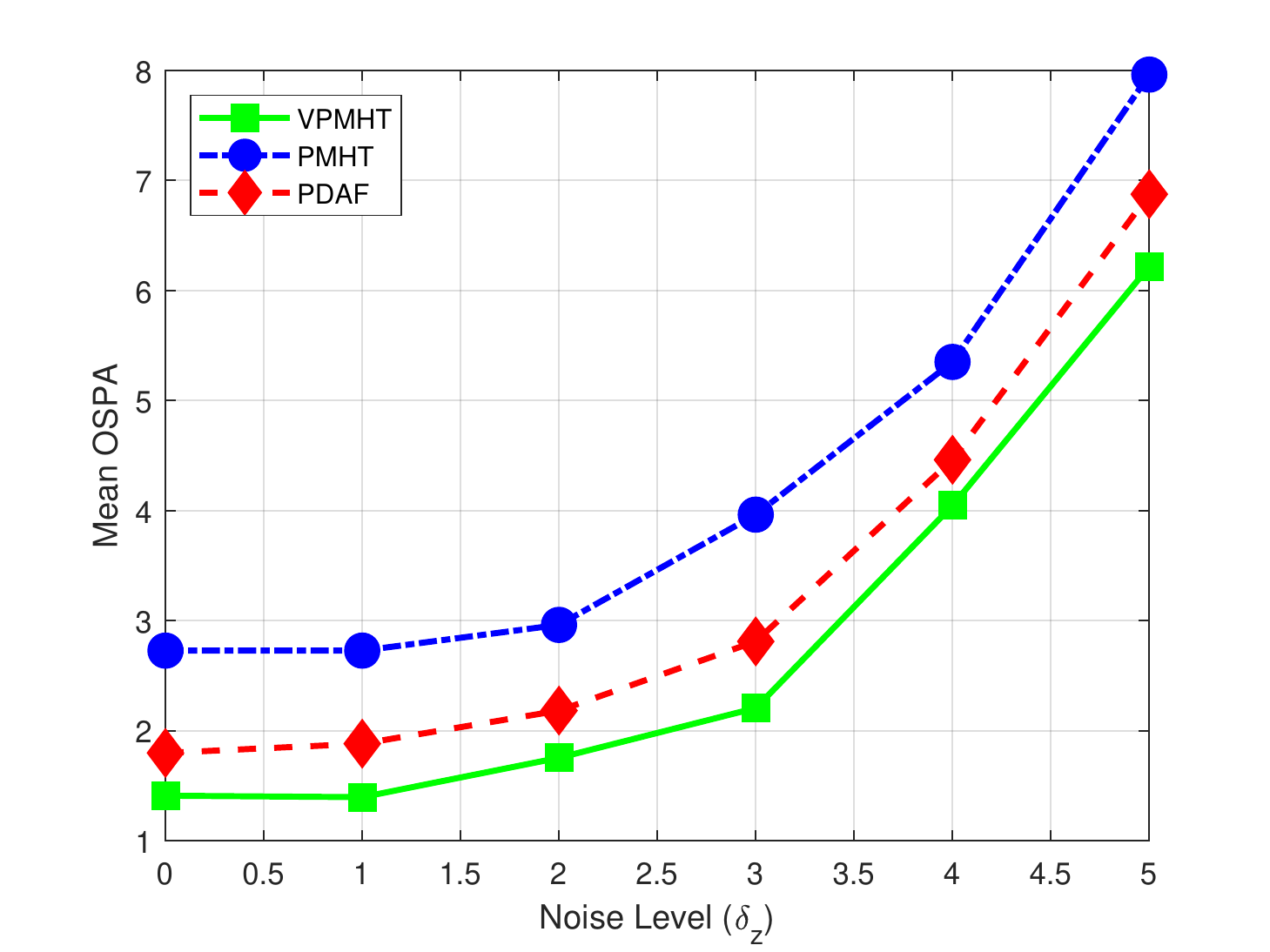}}}
    \caption{Mean OSPA w.r.t different noise level with track-loss}%
    \label{substantial_noise_term}%
\end{figure}

\begin{figure}[H]
    \centering
    \subfloat[$N_{clu} = 0$]{{\includegraphics[scale=.45]{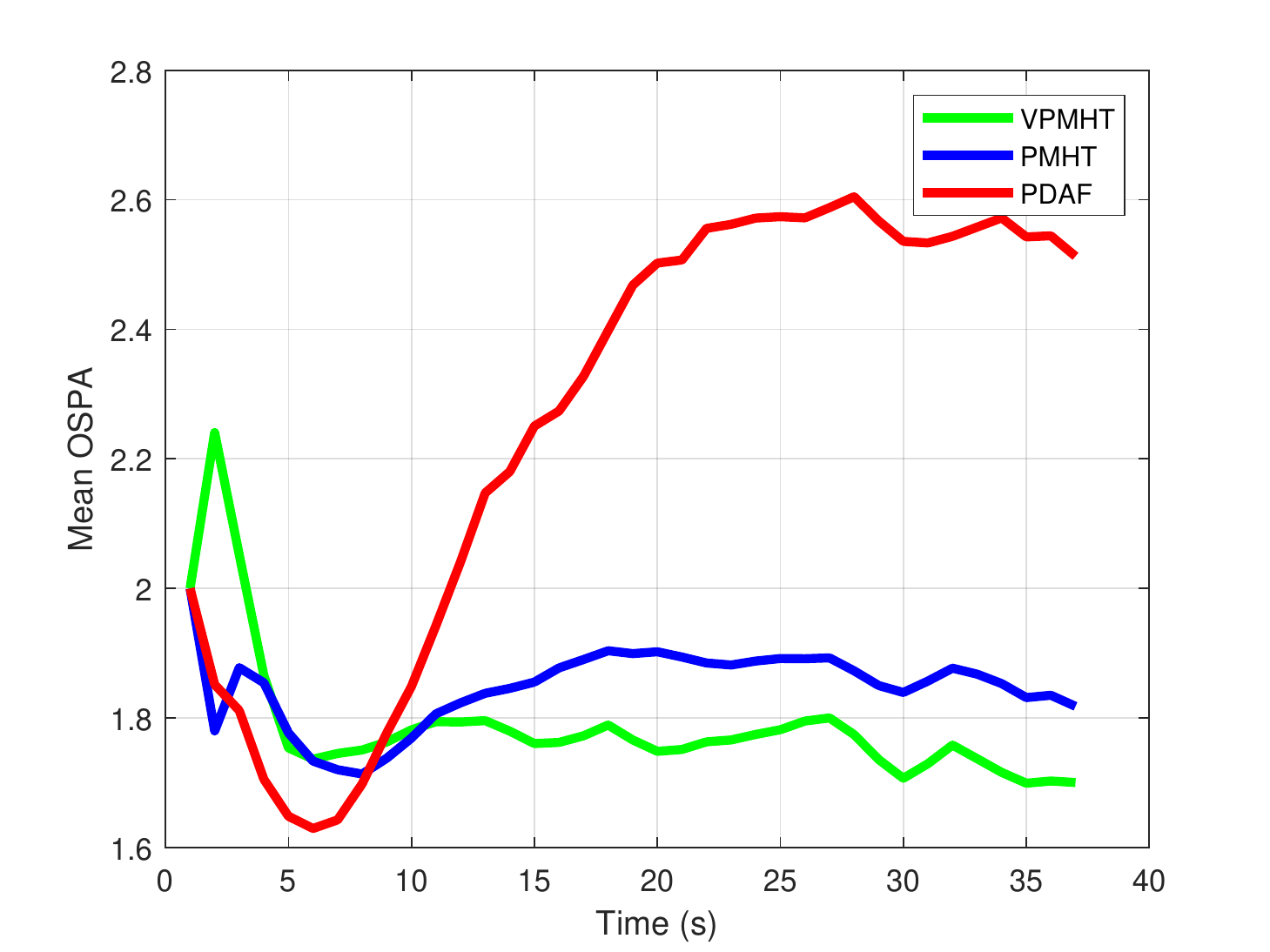}}}
    \subfloat[$N_{clu} = 5$]{{\includegraphics[scale=.45]{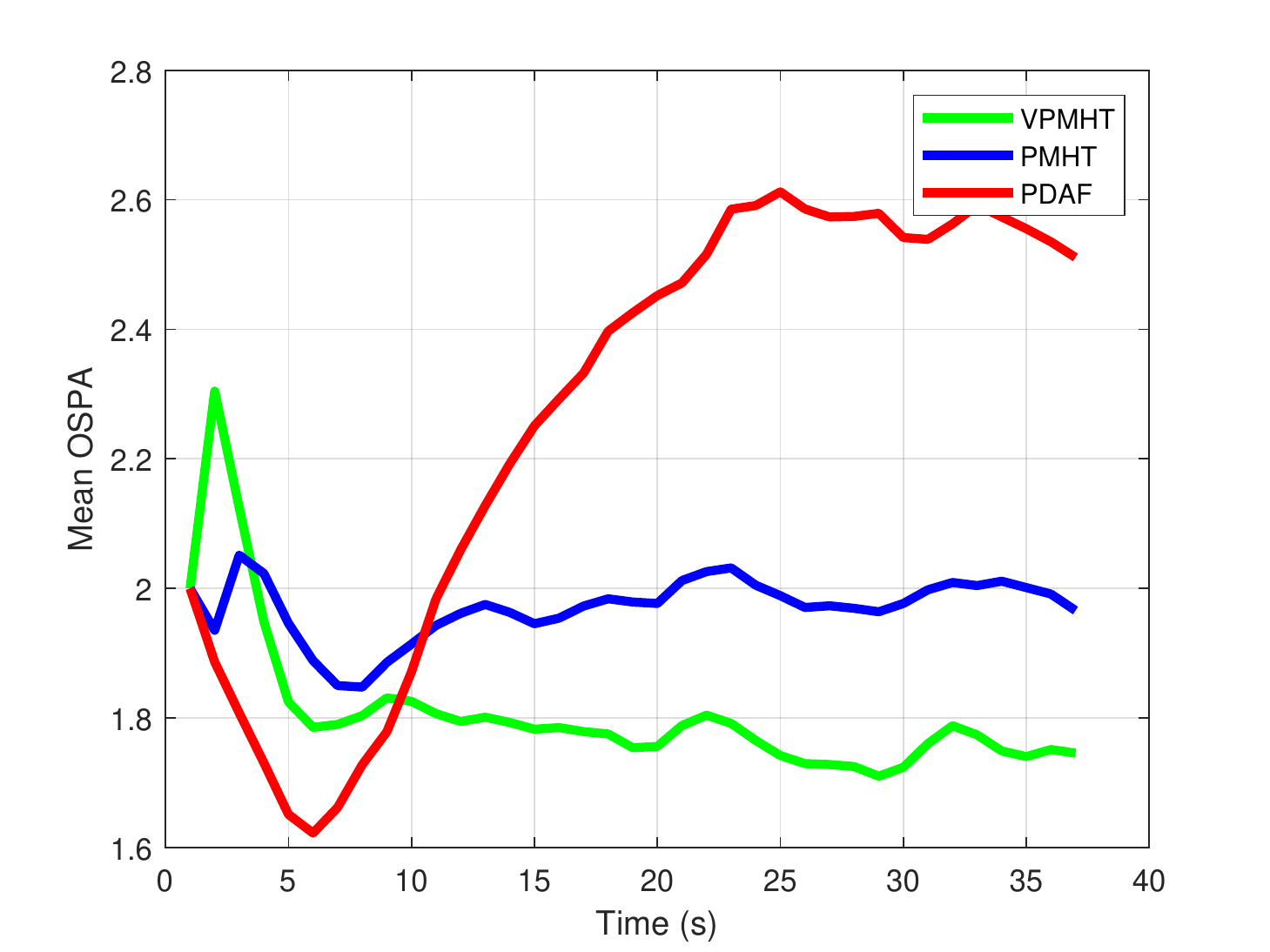}}} \\
    \subfloat[$N_{clu} = 10$]{{\includegraphics[scale=.45]{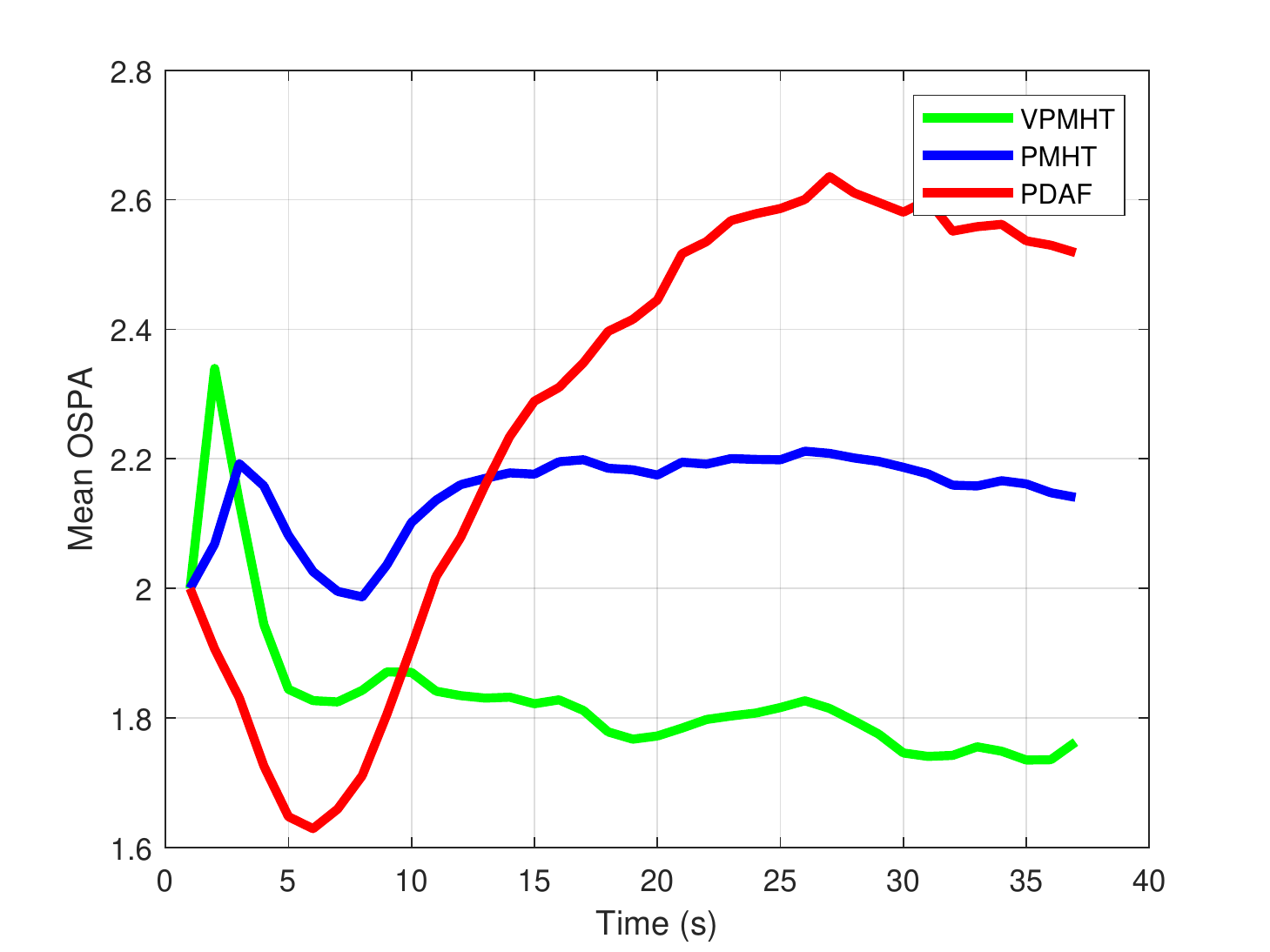}}} 
    \subfloat[$N_{clu} = 15$]{{\includegraphics[scale=.45]{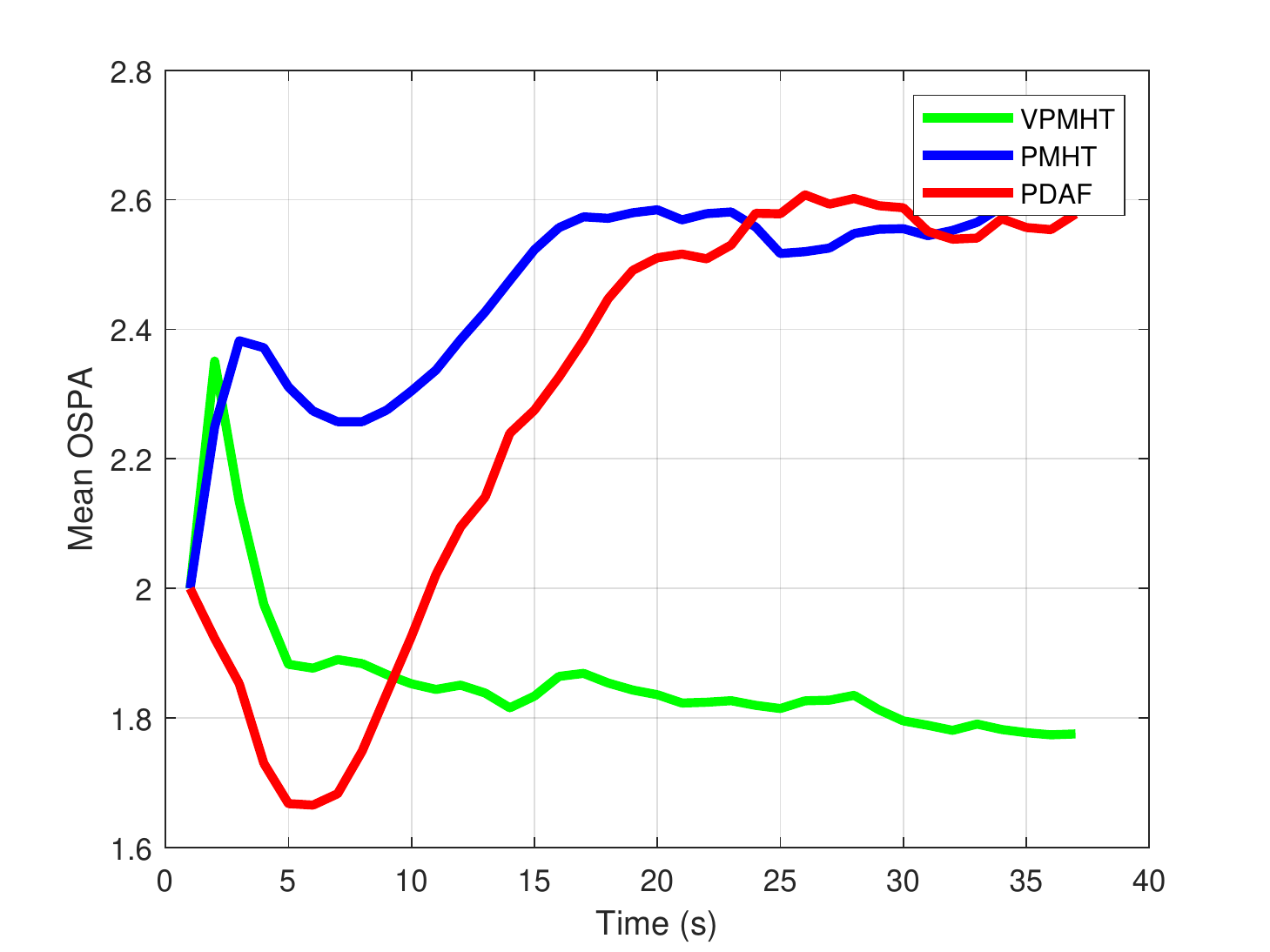}}}\\
    \subfloat[$N_{clu} = 20$]{{\includegraphics[scale=.45]{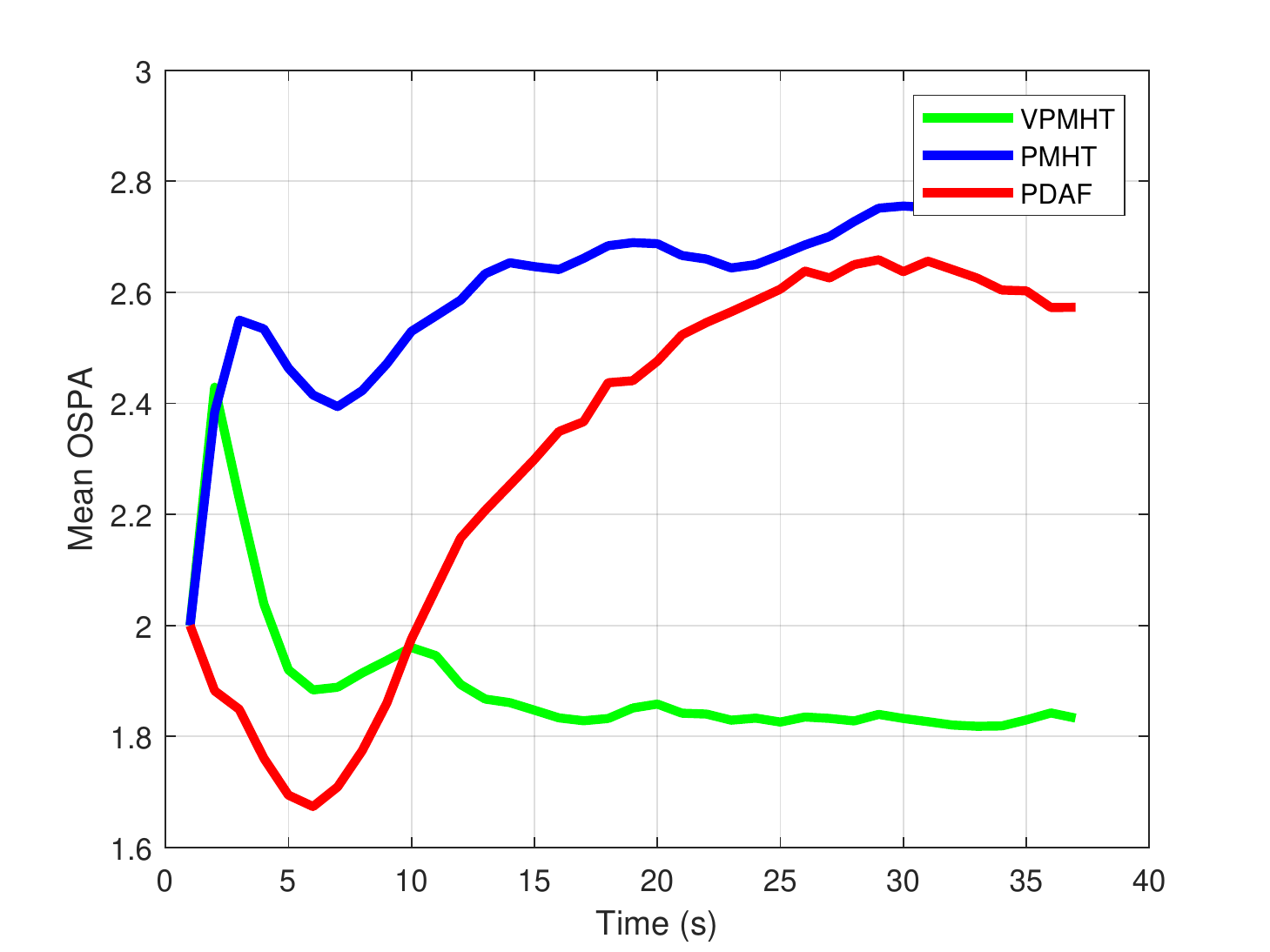}}} 
    \subfloat[Different number of clutters]{{\includegraphics[scale=.45]{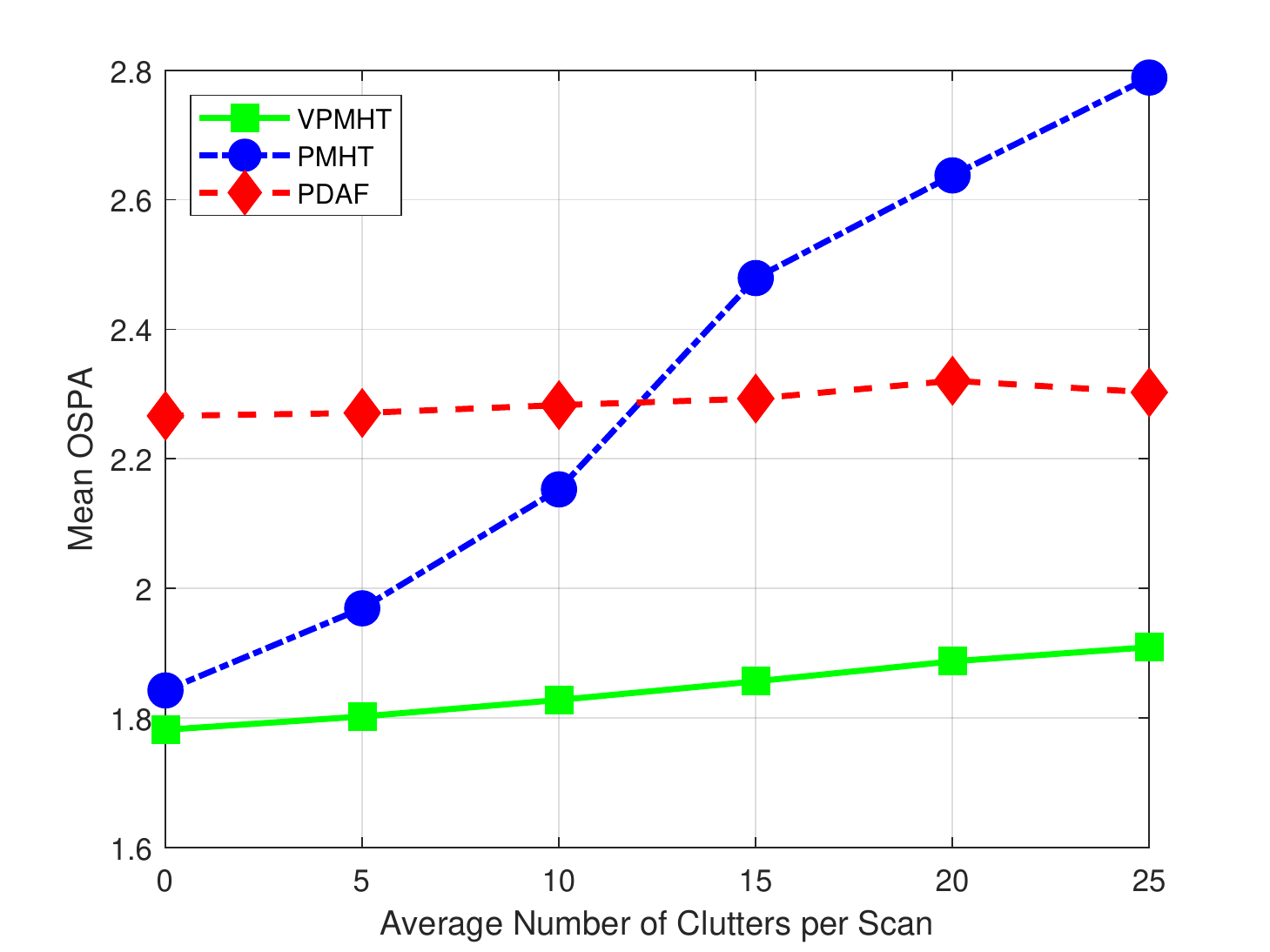}}}
    \caption{Mean OSPA w.r.t to different number of clutters per scan without track-loss}%
    \label{substantial_clutter}%
\end{figure}

\begin{figure}[H]
    \centering
    \subfloat[$N_{clu} = 0$]{{\includegraphics[scale=.45]{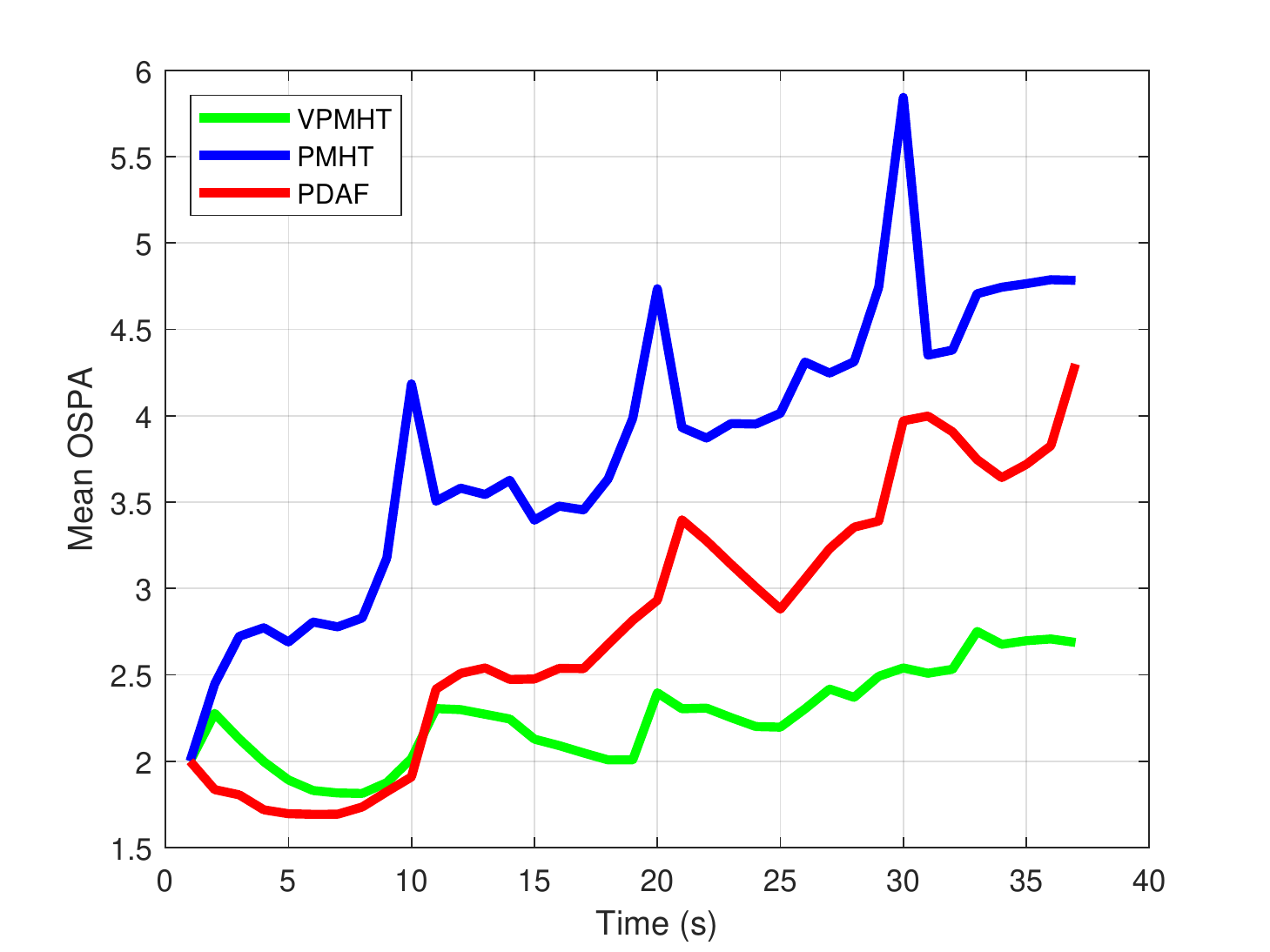}}}
    \subfloat[$N_{clu} = 5$]{{\includegraphics[scale=.45]{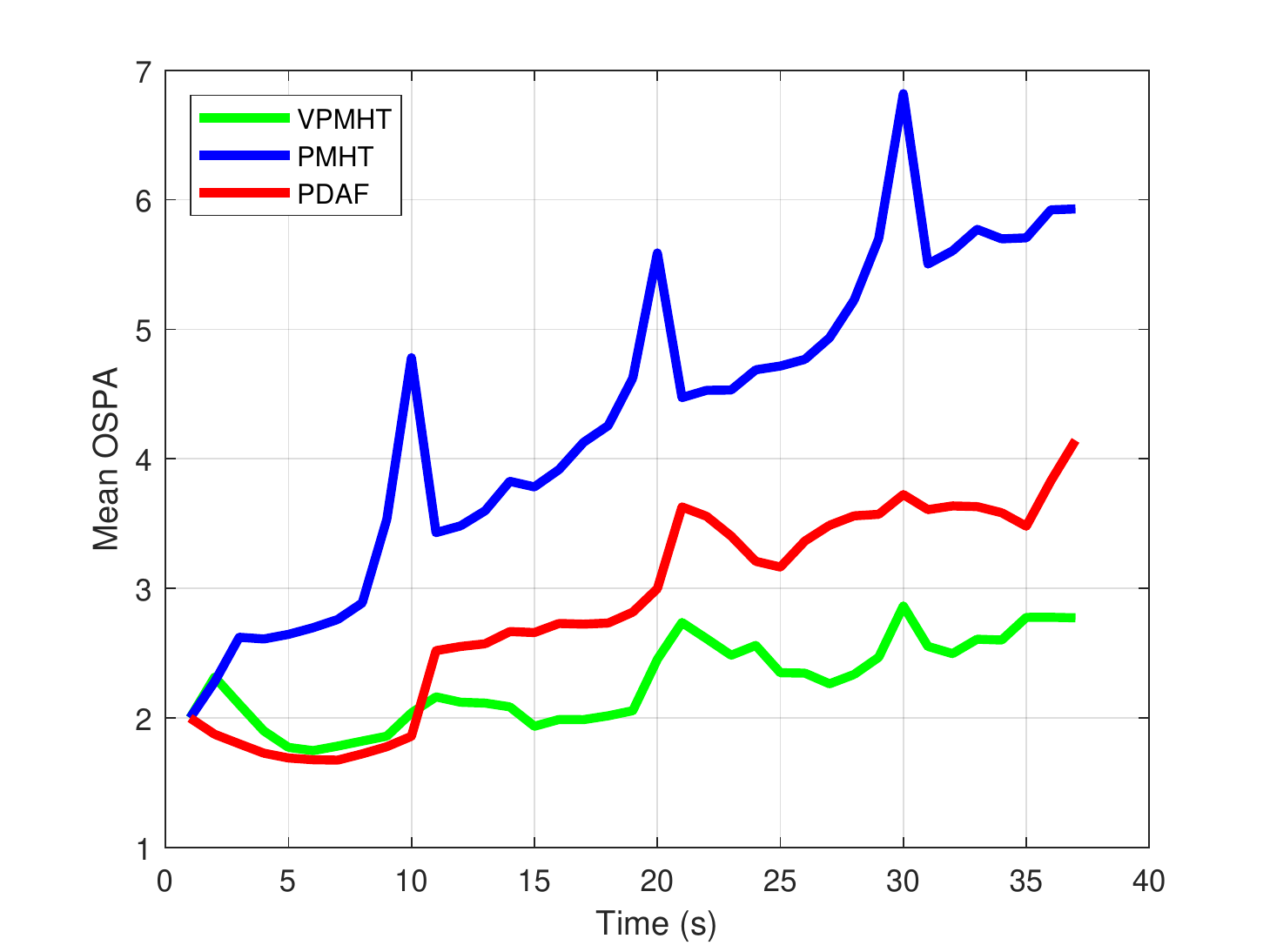}}} \\
    \subfloat[$N_{clu} = 10$]{{\includegraphics[scale=.45]{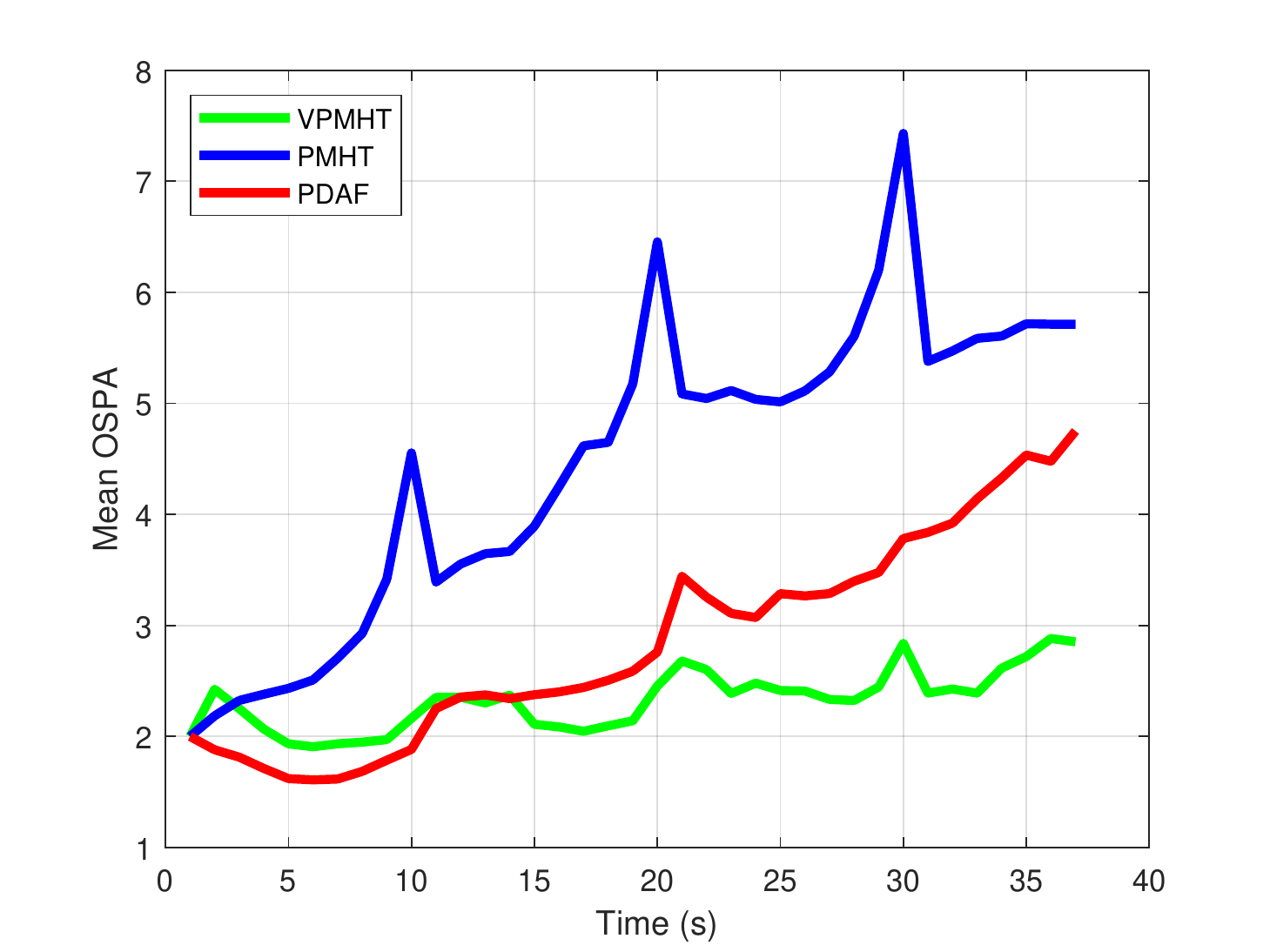}}} 
    \subfloat[$N_{clu} = 15$]{{\includegraphics[scale=.45]{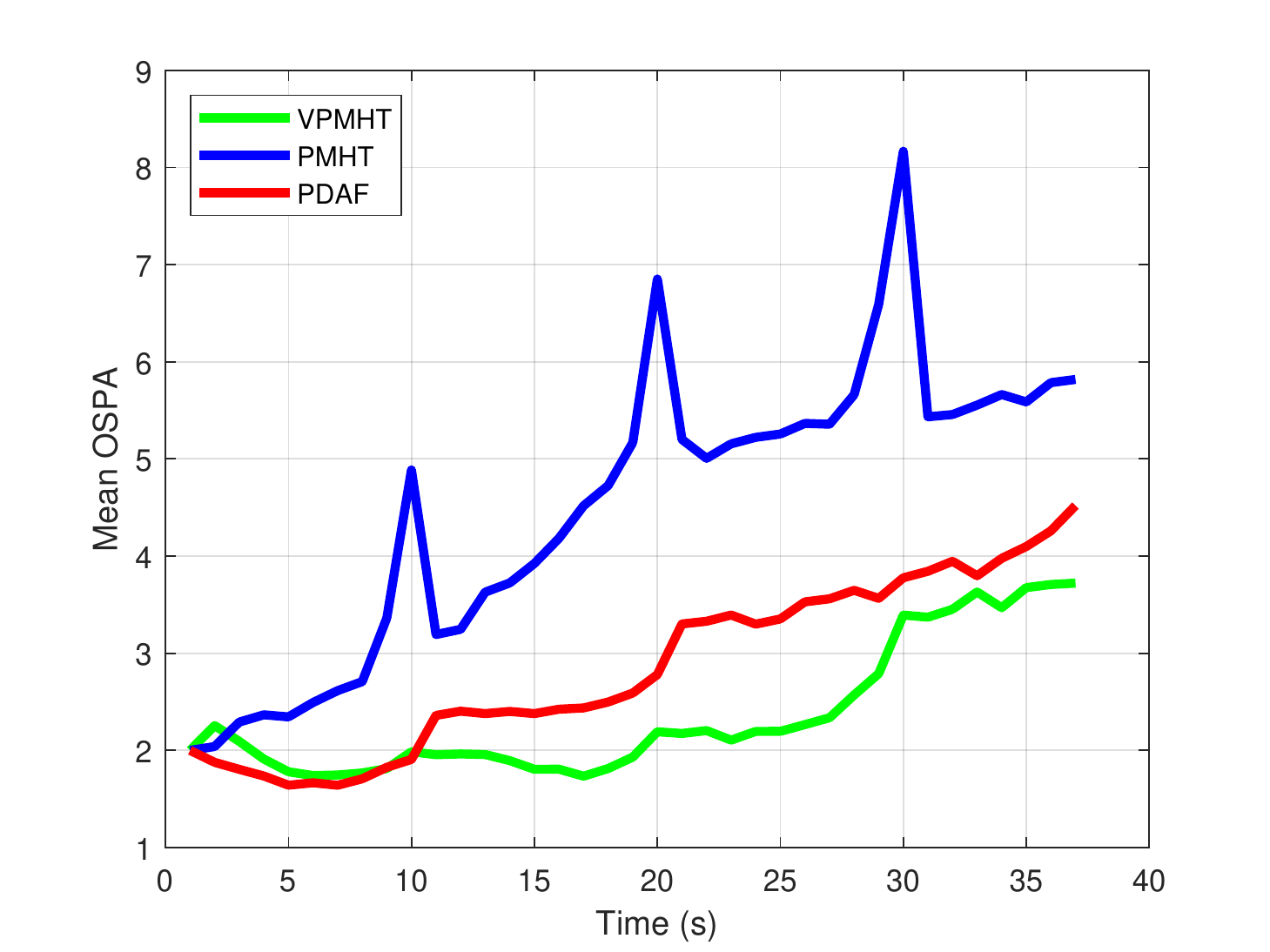}}}\\
    \subfloat[$N_{clu} = 20$]{{\includegraphics[scale=.45]{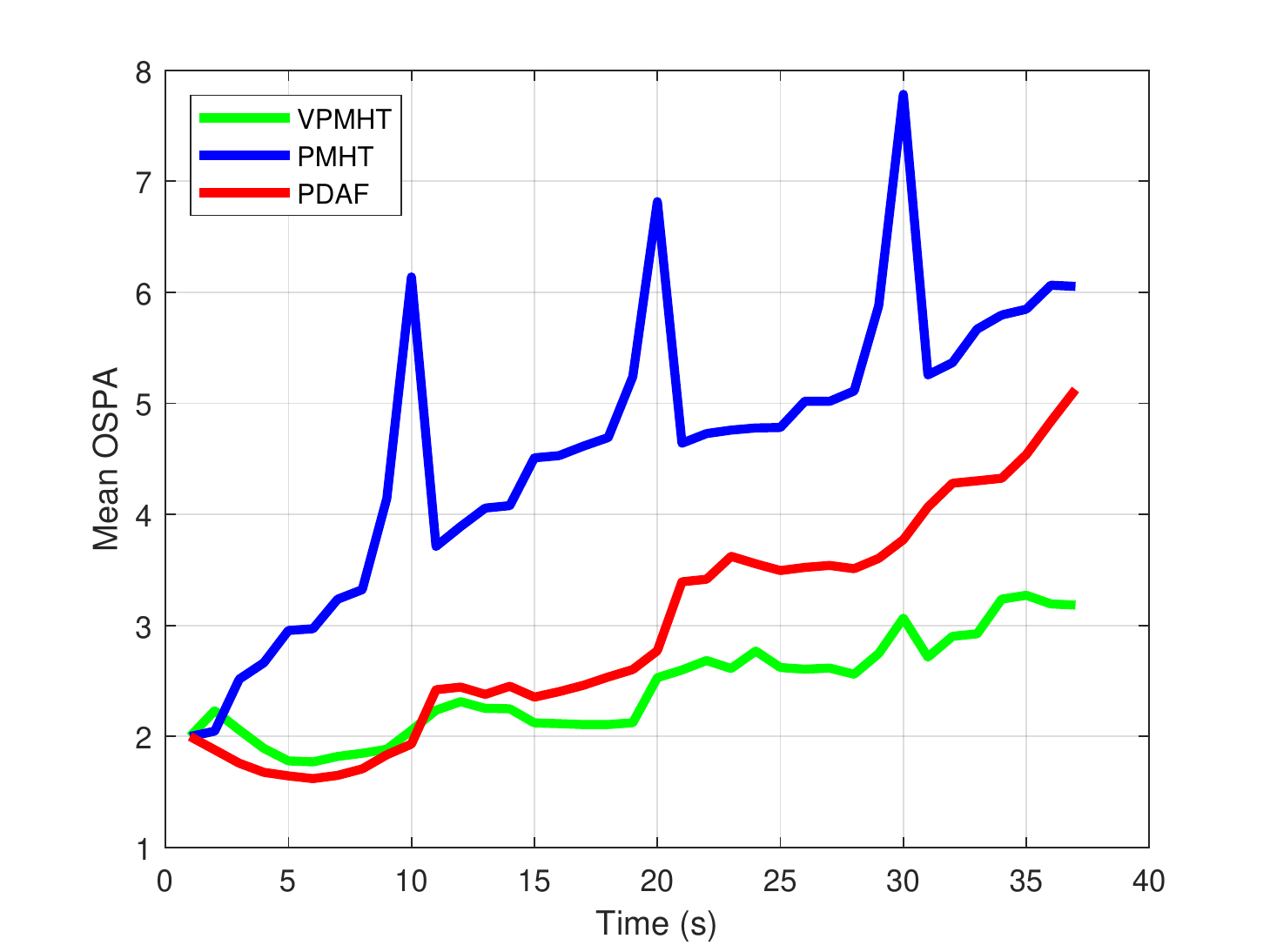}}} 
    \subfloat[Different number of clutters]{{\includegraphics[scale=.45]{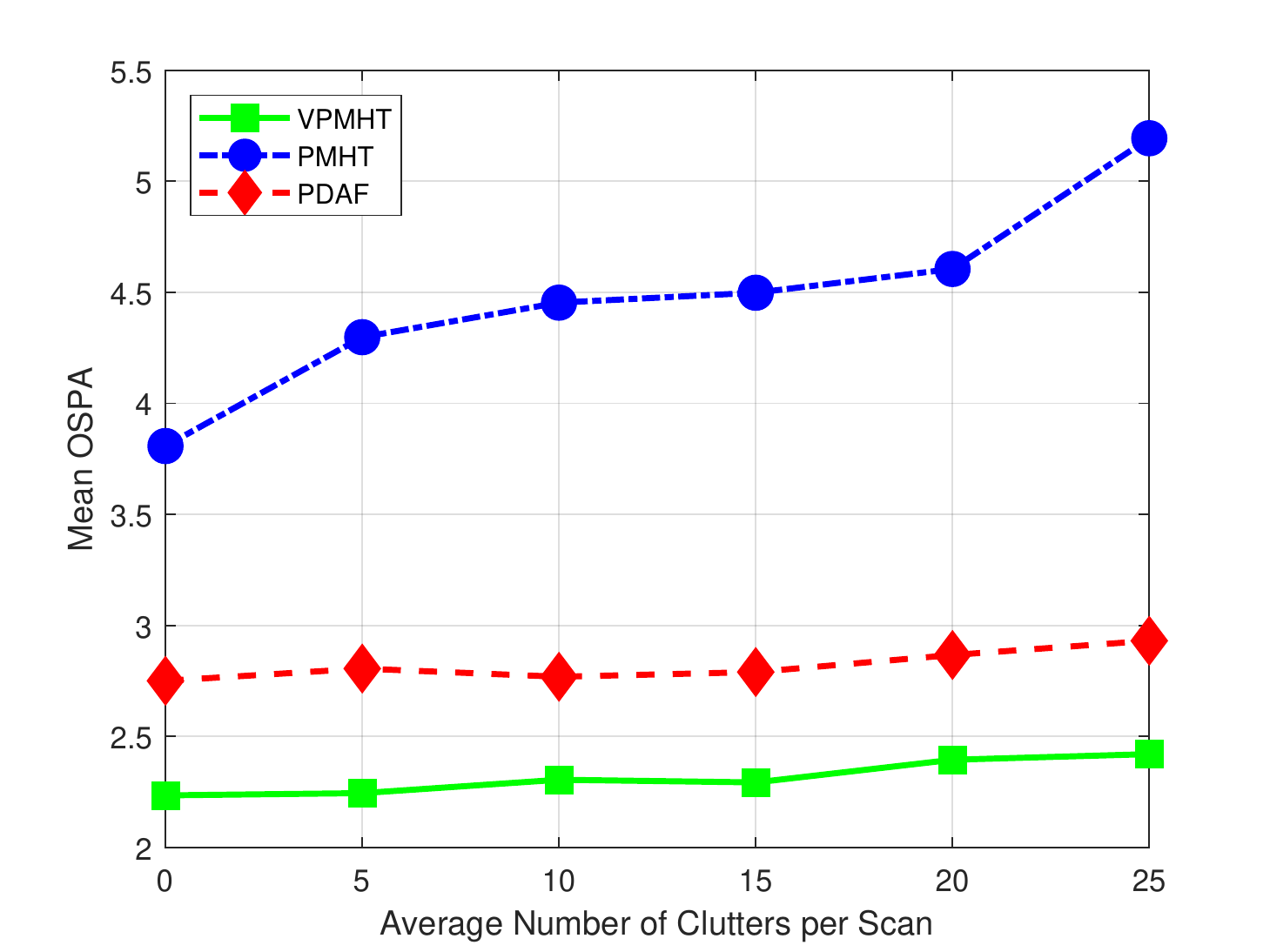}}}
    \caption{Mean OSPA w.r.t different number of clutters per scan with track-loss}%
    \label{substantial_clutter_term}%
\end{figure}

The comparison results among VPMHT, PDAF, and PMHT are presented in Figure. \ref{substantial_noise_vpmht}, \ref{substantial_noise_term}, \ref{substantial_clutter}, and \ref{substantial_clutter_term}, where Figure \ref{substantial_noise_vpmht} and \ref{substantial_noise_term} is for the effect of the measurement noise level and Figure \ref{substantial_clutter} and \ref{substantial_clutter_term} is for the different number of clutter returns. The average run time and convergence iterations at different scenarios for VPMHT, PDAF, and PMHT are presented in Table \ref{mhttab:1}.

\begin{table}[H]
\centering
\caption{Comparison of Computational Efficiency}\label{mhttab:1}
\begin{tabular}{c|c|c|c|c}
\hline\hline
 &  \multicolumn{2}{c}{Time per scan}  & \multicolumn{2}{|c}{Convergence iterations} \\  \hline
 \text{Tracker} &  \text{With track-loss} & \text{No track-loss} & \text{With track-loss} & \text{No track-loss} \\  \hline
\text{PDAF} &  0.0074 & 0.0073 & - & - \\  \hline
\text{PMHT} &  0.0324 & 0.0203 & 22.1155 & 15.3896 \\  \hline
\text{VPMHT} &  0.0451 & 0.0436 & 9.8944 & 9.6894 \\  
\hline\hline
\end{tabular}
\end{table}

cases without track-loss, from Figure \ref{substantial_noise_vpmht} and \ref{substantial_noise_term}, VPMHT shows a moderate tracking accuracy improvement compare to PMHT. This confirms the analysis of the VPMHT that its fusion scheme is more advanced than PMHT. When data association is straightforward, the performance improvement of VPMHT comes from its fusion scheme. As for the comparison with PDAF, our simulation result shows that both PMHT and VPMHT outperforms PDAF. The performance difference between PDAF and the other two algorithms becomes more significant as the level of measurement noise or clutter rate increases. Our simulation result shows the same conclusion as in \cite{willett1998variety, Willett2002pmht} that PMHT outperforms PDAF when there is a high level of measurement noise or clutter rate.

As for the cases with track-loss, see Figure \ref{substantial_clutter} and \ref{substantial_clutter_term}, VPMHT shows a significant performance improvement compares PMHT. It can be identified from the cases with track-loss that the performance improvement of VPMHT among PMHT starts to become more significant as the uncertainty increases. As the noise level or clutter rate grows, the identification of track-loss becomes more challenging since the already lost targets are more likely to get false alarms from nearby tracks or clutters. The effects of clutters are more noticeable than measurement noise since the false alarms are more likely to happen with clutters' growth. The significant performance advantage of VPMHT is expected since the VBEM can handle the model selection problems much better than EM. Our proposed algorithm is even capable of handling the track-loss problem better than PDAF. This might be because a sophisticated fusion scheme will lead to improved data associations when the measurement noise or clutter rate reaches a high level.

It can be identified from Figure \ref{substantial_noise_vpmht}, \ref{substantial_noise_term}, \ref{substantial_clutter}, and \ref{substantial_clutter_term} that the VPMHT outperforms PMHT and PDAF regardless of track-loss. From the comparison of computational efficiency in Table 1, the computation of VPMHT is the slowest. VPMHT is nearly two times slower than the PMHT and six times slower than the PDAF. Both the VPMHT and PDAF shows little fluctuations in computational time when track-losses are introduced. However, PMHT shows a nearly $40\%$ slowdown when track-losses are introduced. This slowdown is lead by the increased iterations for PMHT to reach convergence. It is reasonable to assume that when the identification of track-loss becomes more challenging, the computational burden of VPMHT would be less substantial than PMHT.

It can be summarised from the simulations that:
1) The proposed VPMHT is capable of handling track-loss much better than PMHT and PDAF; 
2) The proposed VPMHT outperforms PMHT and PDAF regardless of track-loss for its advanced fusion scheme; 
3) The computational efficiency of the proposed VPMHT is not as good as PDAF. However, the computational speed of VPMHT approaches PMHT with the introduction of track-loss.

\section{Conclusion}
We have developed a novel MTT algorithm named VPMHT. The proposed VPMHT is based on the modification of the conventional PMHT algorithm. The VBEM implementation of this MTT algorithm is derived. VPMHT is capable of providing a rather accurate tracking result when there are track-losses. Compared to conventional PMHT and PDAF algorithm, the proposed algorithm shows advantages in tracking accuracy, robustness against noisy measurements, background clutter and track-losses. All these advantages of VPMHT come with a non-substantial sacrifice in computational efficiency. Extensive simulations of various scenarios confirm these aspects of the proposed algorithm. As for the future study, it is crucial to deriving an all-in-one framework contains target birth.

\clearpage
\bibliographystyle{elsarticle-num}
\bibliography{IF_VPMHT.bib}

\begin{thebibliography}{10}
\expandafter\ifx\csname url\endcsname\relax
  \def\url#1{\texttt{#1}}\fi
\expandafter\ifx\csname urlprefix\endcsname\relax\def\urlprefix{URL }\fi
\expandafter\ifx\csname href\endcsname\relax
  \def\href#1#2{#2} \def\path#1{#1}\fi

\bibitem{cho2014multi}
H.~Cho, Y.-W. Seo, B.~V. Kumar, R.~R. Rajkumar, A multi-sensor fusion system
  for moving object detection and tracking in urban driving environments, in:
  2014 IEEE International Conference on Robotics and Automation (ICRA), IEEE,
  2014, pp. 1836--1843.

\bibitem{kanade1998advances}
T.~Kanade, R.~Collins, A.~Lipton, P.~Burt, L.~Wixson, Advances in cooperative
  multi-sensor video surveillance, in: Proceedings of DARPA Image Understanding
  Workshop, Vol.~1, Citeseer, 1998, p.~2.

\bibitem{scheunert2004multi}
U.~Scheunert, H.~Cramer, B.~Fardi, G.~Wanielik, Multi sensor based tracking of
  pedestrians: a survey of suitable movement models, in: IEEE Intelligent
  Vehicles Symposium, 2004, IEEE, 2004, pp. 774--778.

\bibitem{konstantinova2003study}
P.~Konstantinova, A.~Udvarev, T.~Semerdjiev, A study of a target tracking
  algorithm using global nearest neighbor approach, in: Proceedings of the
  International Conference on Computer Systems and Technologies, 2003, pp.
  290--295.

\bibitem{blackman2004multiple}
S.~S. Blackman, Multiple hypothesis tracking for multiple target tracking, IEEE
  Aerospace and Electronic Systems Magazine 19~(1) (2004) 5--18.

\bibitem{thomaidis2013multiple}
G.~Thomaidis, M.~Tsogas, P.~Lytrivis, G.~Karaseitanidis, A.~Amditis, Multiple
  hypothesis tracking for data association in vehicular networks, Information
  Fusion 14~(4) (2013) 374--383.

\bibitem{musicki1994integrated}
D.~Musicki, R.~Evans, S.~Stankovic, Integrated probabilistic data association,
  IEEE Transactions on automatic control 39~(6) (1994) 1237--1241.

\bibitem{hamid2015joint}
S.~Hamid~Rezatofighi, A.~Milan, Z.~Zhang, Q.~Shi, A.~Dick, I.~Reid, Joint
  probabilistic data association revisited, in: Proceedings of the IEEE
  international conference on computer vision, 2015, pp. 3047--3055.

\bibitem{he2020distributed}
S.~He, H.-S. Shin, A.~Tsourdos, Distributed multiple model joint probabilistic
  data association with gibbs sampling-aided implementation, Information Fusion
  64 (2020) 20--31.

\bibitem{feldmann2010tracking}
M.~Feldmann, D.~Franken, W.~Koch, Tracking of extended objects and group
  targets using random matrices, IEEE Transactions on Signal Processing 59~(4)
  (2010) 1409--1420.

\bibitem{gilholm2005spatial}
K.~Gilholm, D.~Salmond, Spatial distribution model for tracking extended
  objects, IEE Proceedings-Radar, Sonar and Navigation 152~(5) (2005) 364--371.

\bibitem{baum2011shape}
M.~Baum, U.~D. Hanebeck, Shape tracking of extended objects and group targets
  with star-convex rhms, in: 14th International Conference on Information
  Fusion, IEEE, 2011, pp. 1--8.

\bibitem{zhang2017box}
Y.~Zhang, H.~Ji, Q.~Hu, A box-particle implementation of standard phd filter
  for extended target tracking, Information Fusion 34 (2017) 55--69.

\bibitem{granstrom2019poisson}
K.~Granstr{\"o}m, M.~Fatemi, L.~Svensson, Poisson multi-bernoulli mixture
  conjugate prior for multiple extended target filtering, IEEE Transactions on
  Aerospace and Electronic Systems 56~(1) (2019) 208--225.

\bibitem{he2018multi}
S.~He, H.-S. Shin, A.~Tsourdos, Multi-sensor multi-target tracking using domain
  knowledge and clustering, IEEE Sensors Journal 18~(19) (2018) 8074--8084.

\bibitem{streit1995probabilistic}
R.~L. Streit, T.~E. Luginbuhl, Probabilistic multi-hypothesis tracking, Tech.
  rep., NAVAL UNDERWATER SYSTEMS CENTER NEWPORT RI (1995).

\bibitem{bordonaro2015extracting}
S.~Bordonaro, P.~Willett, Y.~Bar-Shalom, M.~Baum, T.~Luginbuhl, Extracting
  speed, heading and turn-rate measurements from extended objects using the em
  algorithm, in: 2015 IEEE Aerospace Conference, IEEE, 2015, pp. 1--12.

\bibitem{bordonaro2017extended}
S.~Bordonaro, P.~Willett, Y.~Shalom, T.~Luginbuhl, M.~Baum, Extended object
  tracking with exploitation of range rate measurements, ISIF Journal of
  Advances in Information Fusion 12~(2) (2017).

\bibitem{molnar1994application}
K.~Molnar, J.~Modestino, Application of the em algorithm for the
  multitarget/multisensor tracking problem, Signal Processing, IEEE
  Transactions on 46 (1998) 115--129.

\bibitem{krieg1997multisensor}
M.~L. Krieg, D.~A. Gray, Multisensor probabilistic multihypothesis tracking
  using dissimilar sensors, in: Acquisition, Tracking, and Pointing XI, Vol.
  3086, International Society for Optics and Photonics, 1997, pp. 129--138.

\bibitem{logothetis1997maneuvering}
A.~Logothetis, V.~Krishnamurthy, J.~Holst, On maneuvering target tracking via
  the pmht, in: Proceedings of the 36th IEEE Conference on Decision and
  Control, Vol.~5, IEEE, 1997, pp. 5024--5029.

\bibitem{ruan1998pmht}
Y.~Ruan, P.~Willett, R.~Streit, The pmht for maneuvering targets, in:
  Proceedings of the 1998 American Control Conference. ACC (IEEE Cat. No.
  98CH36207), Vol.~4, IEEE, 1998, pp. 2432--2433.

\bibitem{Willett2002pmht}
P.~Willett, Y.~Ruan, R.~Streit, Pmht: Problems and some solutions, IEEE
  Transactions on Aerospace and Electronic Systems 38~(3) (2002) 738--754.

\bibitem{dunham2002hybrid}
D.~T. Dunham, R.~G. Hutchins, Hybrid tracking algorithm using mht and pmht, in:
  Signal and Data Processing of Small Targets 2002, Vol. 4728, International
  Society for Optics and Photonics, 2002, pp. 166--175.

\bibitem{wieneke2007sequential}
M.~Wieneke, W.~Koch, On sequential track extraction within the pmht framework,
  EURASIP Journal on Advances in Signal Processing 2008 (2007) 1--13.

\bibitem{wieneke2008track}
M.~Wieneke, P.~Willett, On track-management within the pmht framework, in: 2008
  11th International Conference on Information Fusion, IEEE, 2008, pp. 1--8.

\bibitem{rago1995comparison}
C.~Rago, P.~Willett, R.~Streit, A comparison of the jpdaf and pmht tracking
  algorithms, in: 1995 International conference on acoustics, speech, and
  signal processing, Vol.~5, IEEE, 1995, pp. 3571--3574.

\bibitem{rago1995direct}
C.~{Rago}, P.~{Willett}, R.~{Streit}, Direct data fusion using the pmht, in:
  Proceedings of 1995 American Control Conference - ACC'95, Vol.~3, 1995, pp.
  1698--1702 vol.3.
\newblock \href {https://doi.org/10.1109/ACC.1995.529798}
  {\path{doi:10.1109/ACC.1995.529798}}.

\bibitem{willett1998variety}
P.~Willett, Y.~Ruan, R.~Streit, A variety of pmhts, in: Proceedings of the
  workshop Probabilistic methods in multi-target tracking, ENST, Paris, 1998.

\bibitem{jeong2002based}
H.~Jeong, J.-H. Park, An em-based adaptive multiple target tracking filter,
  International Journal of Adaptive Control and Signal Processing 16~(1) (2002)
  1--23.

\bibitem{rago1995modified}
C.~Rago, P.~Willett, R.~Streit, A modified pmht, in: Proceedings of the 1995
  Conference on Information Sciences and Systems, 1995.

\bibitem{lan2016survey}
H.~Lan, X.~Wang, Q.~Pan, F.~Yang, Z.~Wang, Y.~Liang, A survey on joint tracking
  using expectation--maximization based techniques, Information Fusion 30
  (2016) 52--68.

\bibitem{lundgren2015variational}
M.~Lundgren, L.~Svensson, L.~Hammarstrand, Variational bayesian expectation
  maximization for radar map estimation, IEEE Transactions on Signal Processing
  64~(6) (2015) 1391--1404.

\bibitem{Approximate_Inference}
C.~M. Bishop, Pattern recognition and machine learning, springer, 2006, Ch.~10,
  pp. 461--521.

\bibitem{bilmes1998gentle}
J.~A. Bilmes, et~al., A gentle tutorial of the em algorithm and its application
  to parameter estimation for gaussian mixture and hidden markov models,
  International Computer Science Institute 4~(510) (1998) 126.

\bibitem{he2020information}
S.~He, H.-S. Shin, A.~Tsourdos, Information-theoretic joint probabilistic data
  association filter, IEEE Transactions on Automatic Control (2020).

\bibitem{kalman1960new}
R.~E. Kalman, A new approach to linear filtering and prediction problems
  (1960).

\bibitem{rauch1965maximum}
H.~E. Rauch, F.~Tung, C.~T. Striebel, Maximum likelihood estimates of linear
  dynamic systems, AIAA journal 3~(8) (1965) 1445--1450.

\bibitem{springer2013mathematical}
T.~V. Springer, Mathematical analysis and computational methods for
  probabilistic multi-hypothesis tracking (pmht), Ph.D. thesis, Universit{\"a}t
  Ulm (2013).

\bibitem{hogg2005introduction}
R.~V. Hogg, J.~McKean, A.~T. Craig, Introduction to mathematical statistics,
  Pearson Education, 2005.

\bibitem{kay1993fundamentals}
S.~M. Kay, Fundamentals of statistical signal processing, Prentice Hall PTR,
  1993.

\bibitem{schuhmacher2008consistent}
D.~Schuhmacher, B.-T. Vo, B.-N. Vo, A consistent metric for performance
  evaluation of multi-object filters, IEEE transactions on signal processing
  56~(8) (2008) 3447--3457.

\end{thebibliography}

\end{document}